\RequirePackage{etoolbox}
\documentclass[12pts]{imsart}

\pubyear{0000}
\volume{00}
\issue{0}
\doi{0000}
\firstpage{1}
\lastpage{1}

\usepackage{amsthm}
\usepackage{amsmath}
\usepackage{natbib}
\usepackage[colorlinks,citecolor=blue,urlcolor=blue,filecolor=blue,backref=page]{hyperref}
\usepackage{graphicx}
\usepackage{booktabs}
\usepackage{multirow}
\usepackage{subfigure,float}
\usepackage{arydshln}
\usepackage[normalem]{ulem}
\usepackage[usenames, dvipsnames]{color}

\newcommand{\be}{\begin{equation}}
\newcommand{\ee}{\end{equation}}
\newcommand{\bea}{\begin{eqnarray}}
\newcommand{\eea}{\end{eqnarray}}
\newcommand{\mbf}[1]{\mathbf{#1}}
\newcommand{\mbs}[1]{\boldsymbol{#1}}

\renewcommand{\d}{\text{d}}
\newcommand{\by}{\mbf{y}}

\newcommand{\bx}{\mbf{x}}
\newcommand{\bq}{\mbf{q}}
\newcommand{\bz}{\mbf{z}}
\newcommand{\bv}{\mbf{v}}

\newcommand{\bw}{\mbf{w}}

\newcommand{\bX}{\mbf{X}}
\newcommand{\bQ}{\mbf{Q}}

\newcommand{\bmu}{\mbs{\mu}}

\newcommand{\blambda}{{\mbs{\lambda}}}
\newcommand{\balpha}{{\mbs{\alpha}}}

\newcommand{\btheta}{{\mbs{\theta}}}

\newcommand{\bI}{\mbf{I}}
\newcommand{\bH}{\mbf{H}}

\newcommand{\1}{\mbs{1}}
\newcommand{\0}{\mbs{0}}

\renewcommand{\exp}[1]{\text{exp}\left[#1\right]}
\newcommand{\M}{{M}}

\newcommand{\MF}{{M_F}}
\newcommand{\MFsub}[1]{{M_{F_{#1}}}}

\newcommand{\cM}{\mathcal{M}}

\newcommand{\lrp}[1]{\left(#1\right)}
\newcommand{\lrb}[1]{\left\{#1\right\}}
\newcommand{\lrsqb}[1]{\left[#1\right]}

\renewcommand{\d}{\text{d}}

\newcommand{\bean}{\begin{eqnarray*}}
\newcommand{\eean}{\end{eqnarray*}}

\newcommand{\jd}{J_\bullet}
\startlocaldefs
\numberwithin{equation}{section}
\theoremstyle{plain}

\endlocaldefs

\begin{document}

\begin{frontmatter}
\title{Intrinsic Bayesian Analysis for  Occupancy Models}
\runtitle{Objective Bayesian Selection for Occupancy Models}

\begin{aug}
\author{\fnms{Daniel} \snm{Taylor-Rodr\'{i}guez}\thanksref{addr1}\ead[label=e1]{dt108@stat.duke.edu}},
\author{\fnms{Andrew J.} \snm{Womack}\thanksref{addr2}\ead[label=e2]{ajwomack@indiana.edu}}
\author{\fnms{Claudio} \snm{Fuentes}\thanksref{addr3}\ead[label=e3]{fuentesc@stat.oregonstate.edu}}
\and
\author{\fnms{Nikolay} \snm{Bliznyuk}\thanksref{addr4}\ead[label=e4]{nbliznyuk@ufl.edu}}

\runauthor{Taylor-Rodriguez et al.}

\address[addr1]{First coauthor, Postdoctoral Fellow, SAMSI/Duke University, Research Triangle Park, NC 27709
    \printead{e1} % print email address of "e1"
   % \printead*{e2}
}

\address[addr2]{First coauthor, Assistant Professor, 
Department of Statistics, 
Indiana University, 
Bloomington, IN 47408. 
    \printead{e2}
  %  \printead{u1}
}

\address[addr3]{Assistant Professor, 
Department of Statistics, 
Oregon State University, 
Corvallis, OR  97331.
    \printead{e3}
}

\address[addr4]{Corresponding author. Assistant Professor, 
Departments of Agricultural and Biological Engineering, Biostatistics and Statistics,
University of Florida,
Gainesville, FL 32611.
    \printead{e4}
  % \printead{u1}
}

\end{aug}

%----------------------------------------
%\author{Daniel Taylor-Rodr\'{i}guez \thanks{Postdoctoral Fellow, SAMSI/Duke University, Research Triangle Park, NC 27709. Email:  \texttt{taylor-rodriguez@samsi.info}}\\
%SAMSI/Duke University
%\and 
%Nikolay Bliznyuk\thanks{Assistant Professor, Department of Agricultural and Biological Engineering, University of Florida,
%Gainesville, FL 32611. Email:  \texttt{nbliznyuk@ufl.edu}} \\
%University of Florida
%\and
%Andrew J. Womack\thanks{Assistant Professor, Department of Statistics, Indiana University, Bloomington, IN 47408. Email: \texttt{ajwomack@indiana.edu}}\\
%Indiana University
%\and
%Claudio Fuentes\thanks{Assistant Professor, Department of of Statistics, Oregon State University, Corvallis, OR  97331.  Email:  \texttt{fuentesc@stat.oregonstate.edu}}\\
%Oregon State University 
%}
%\date{\today}
%\maketitle
%----------------------------------------

\begin{abstract}
Occupancy models are typically used to determine the probability of a species being present at a given site while accounting for imperfect detection.  The survey data underlying these models often include information on several predictors that could potentially characterize habitat suitability and species detectability. Because these variables might not all be relevant, model selection techniques are necessary in this context. In practice, model selection is performed using the Akaike Information Criterion (AIC), as few other alternatives are available.   This paper builds an objective Bayesian variable selection framework for occupancy models through the intrinsic prior methodology. The procedure incorporates priors on the model space that account for test multiplicity and respect the polynomial hierarchy of the predictors when higher-order terms are considered.   The methodology is implemented using a stochastic search algorithm that is able to thoroughly explore large spaces of occupancy models.  The proposed strategy is entirely automatic and provides control of false positives without sacrificing the discovery of truly meaningful covariates. The performance of the method is evaluated and compared to AIC through a simulation study. The method is illustrated on two datasets previously studied in the literature.
\end{abstract}



\end{frontmatter}

\section{Introduction}\label{sec:intro}
It is often the case that measurements recorded for a given response are, at best, a noisy version of the variable of interest. A particular case of this issue is known as imperfect detection, and constitutes a pervasive problem.  For instance, in biological surveys subject to imperfect detection, ``presence/absence'' data for a given species  actually  become ``detection/non-detection'' data, since a specie may be present at a given site, but may go undetected in a survey. Ignoring imperfect detection may lead to inaccurate measurement of the presences \citep{Guillera-Arroita2014}, which generally results in  biased parameter estimates \citep{MacKetal_02}.  

As defined in the ecological literature, \emph{occupancy} is the proportion of sites where a target species is present, constituting a state variable instrumental to assess the distribution of species \citep{MacKetal_02}.  Over the past decade, site occupancy models have been the main tool used by ecologists to estimate occupancy while accounting for imperfect detection.  Occupancy models describe the observed data by linking two processes: presence  and detection. Occupancy models adapt the conventional binary regression model to produce separate estimates for presence and detection probabilities \citep{DTR_12}.  This separation is possible by surveying repeatedly the sampling locations, which provides additional information to better assess if non-detection of the specie truly corresponds to its absence. Conveniently, these models can be fitted even if the number of surveys is unbalanced across sites.  The core of the occupancy model is characterized by the hierarchy
\bea
y_{ij}|z_i&\sim&\text{Bern}(z_i p_{ij})\nonumber\\
z_{i}&\sim&\text{Bern}(\psi_{i}),
\label{eq:occbase}
\eea
where $y_{ij}$ is the binary detection indicator at the $i$th site ($i=1,\ldots,N$) during the $j$th survey ($j=1,\ldots,J_i$). The detection probability  for event $\lrb{y_{ij}=1}$ is $p_{ij}$ whenever the species is present; and $z_{i}$ is the presence indicator at the $i$th site with success probability $\psi_i$. Note that the $z_i$ are imperfectly observed. At at site $i$, whenever the vector of detections $\by_i\neq \boldsymbol{0}$, then we know that $z_i=1$, but $\by_i=\boldsymbol{0}$ does not imply that $z_i=0$. To produce estimates of $\psi_i$ and $p_{ij}$, site occupancy surveys collect information on several predictors with the potential to influence habitat suitability (characterizing $\psi_i$) and species detectability (defining $p_{ij}$).  Given that some of the  collected predictors may be uninformative or redundant, variable selection techniques are instrumental in identifying good models.  

In this paper, we propose  an objective Bayesian variable selection procedure for occupancy models. Our approach is based on intrinsic objective priors for the model parameters, and uses priors over the model space that simultaneously account for test multiplicity, and, if interactions and/or polynomial terms are considered, enforces the polynomial hierarchical structure among predictors.

Currently, variable selection procedures for occupancy models implemented in statistical software are mainly based on the Akaike Information Criterion (AIC) \citep{A_73}.   As a consequence, these procedures do  not allow for valid post-selection inference and uncertainty quantification, and typically require enumerating and fitting every  possible model in the space of models under consideration \citep[e.g.,][]{Mazerolle2013,FiskeUNM}. In practice, this enumeration is feasible  only if the model space is small enough, either because substantial knowledge about the underlying ecological processes is available to constrain the model space, or because only a few variables are considered to begin with. Nevertheless, many site occupancy surveys collect large amounts of covariate information about the sampled sites,  and since the total number of candidate models grows exponentially in the number of predictors, choosing a reduced set of models based on ecological intuition becomes increasingly difficult.  

The AIC is designed to find the model that is the closest to the true (unknown) model with respect to Kullback-Leibler divergence, identifying as good models those with smaller AIC values. It has been shown, however, that the AIC has certain limitations as a model selection criterion. For instance, if nested models are being considered, the AIC will not necessarily select the true model \citep{Wasserman2000}. In fact, the AIC generally shows a weak signal-to-noise ratio and  tends to prefer more complex models, even if the true model is available \citep{rao2001}.  Other versions of the AIC address this issue by including a bias correction factor that enhances the signal-to-noise ratio \citep[see][]{Hurvich1989,McQuarrie1997}; however, these modified versions cannot be used with occupancy models, as they depend on the effective sample size, which is unknown for these models.

In this context, Bayesian methods are more appealing. Under regularity conditions, when the true model is contained in a fixed model space, its posterior probability converges to one as the number of sites and surveys per site both increase. In addition, if the true model is not contained in the model space, the posterior probability of the most parsimonious model closest to the true data generating process tends asymptotically to one. In the finite sample context, Bayesian methods allow for full and faithful error propagation. Furthermore, the Bayesian machinery provides the means to conduct valid inference accounting for model uncertainty.

A Bayesian selection procedure for occupancy models was described in \citet{Hooten2015}. However,  their implementation uses informative prior distributions on the model parameters, tailored specifically to the example discussed in the paper, which prevents the approach from being applicable to occupancy problems in general.  It is often the case that subjective elicitation of parameter and model prior distributions is not possible, since neither the relationship between the response and the predictors, nor the advantages of one model over another, are clearly understood. In addition, the use of seemingly innocuous subjective priors may drastically affect outcomes.  This has been a recurring argument in favor of objective Bayesian procedures, which appeal to the use of formal rules to build parameter priors that incorporate the structural information inside the likelihood while utilizing some objective criterion \citep{Kass1996}.  

To the best of our knowledge, the method proposed in this article is the first general Bayesian selection procedure for occupancy models, that
\begin{enumerate}
\item  bypasses the need for hyper-parameter tuning,
\item  uses priors specifically designed for  testing,
\item  controls for test multiplicity, and
\item  accounts for the hierarchical polynomial structure in the predictors.
\end{enumerate}

In building our approach, we first derive intrinsic priors \citep{Berger1996,MBR_98} for the model parameters in both the presence and detection components of the single-season occupancy model.  For the model priors, we consider the ones proposed in \citet{Taylor-Rodriguez2015}.  These priors, in addition to controlling for test multiplicity, allow restricting the model space to the set of models that respect (weakly or strongly) the polynomial hierarchy among the predictors whenever interactions and higher-order terms are considered. As discussed in \citet{P_87, P_90} when covariate interactions and polynomial terms are present, failure to restrict the class of models  to those respecting strong heredity may result in incoherent variable selection. This is because the model design matrices are not invariant to linear transformations of order-one predictors (e.g., recentering of the main effect variables). Using the derived intrinsic priors on the parameter space and the multiplicity correction priors on the model space, we build a fast stochastic search algorithm that allows us to thoroughly explore large spaces for the single-season occupancy model framework.  This strategy is completely automatic, avoiding the need for both tuning parameters in the sampling algorithm and subjective elicitation of parameter prior distributions. Furthermore, as any other Bayesian approach, it naturally enables parameter and model uncertainty quantification.

The outline of the paper is as follows: in Section \ref{sec:prelims}, we provide  background on occupancy models and set notation. In Section \ref{subsec:notation}, we introduce our objective Bayesian model selection method and develop the Gibbs sampler. In Section \ref{sec:sims}, we present results from a simulation study and a comparison with selection using AIC. In Section \ref{subsec:bluehawk}, we illustrate our methodology on two datasets, which have been previously examined in the ecological literature \citep{Kery2005,KG_10,DTR_12}. We conclude the paper with a brief discussion. Code for all the tools proposed is available in the R package \textsf{OccOBayes}.  A description of the stochastic search algorithm is included in the Appendix.

\section{Inference for a single model}\label{sec:prelims}

This section briefly describes the estimation procedure for a single model.  Assuming the probit link, the occupancy model can be characterized in terms of latent variables, which in turn allows one to relate the detection and occupancy probabilities to predictors. We build an objective prior distribution for the regression coefficients using the expected posterior prior framework \citep{Perez2002} where we condition on both the observed data as well as the unobserved latent variables \citep{Leon-Novelo2012}.

\subsection{The Occupancy Model with Probit Link}\label{subsec:occmod}

The occupancy model in \eqref{eq:occbase} is completed in two ways. First, the probabilities for detection $p_{ij}$ and for presence $\psi_i$ are linked to vectors of predictors $\bq_{ij}$ and $\bx_i$, respectively, through appropriate link functions, $g_p(p_{ij})=\bq_{ij}^\prime \blambda$ and $g_\psi(\psi_i)=\bx_i^\prime\balpha$. We assume that the link function is the inverse standard normal cdf, leading to probit models. Other binary regression models can be fit and lead to slightly more complicated computational algorithms. Second, the parameters of the underlying space, here $(\balpha, \blambda)$, are given a prior distribution $\pi(\balpha,\blambda)$. This paper proposes a prior distribution building on the expected posterior prior method of \citet{Leon-Novelo2012}.

Letting $\bX$ and $\bQ$ be the matrices whose rows are, respectively, vectors $\bx_i^\prime$ and $\bq_{ij}^\prime$ for $i=1,\ldots,N$ and $j=1,\ldots,J_i$, the Bayesian probit occupancy model is specified as
\bea
y_{ij}|z_i,\balpha,\blambda,\bQ,\bX&\sim&\text{Bern}(z_i p_{ij})\quad\text{with}\quad
p_{ij}=\Phi\left(\bq_{ij}^\prime \blambda\right)\nonumber\\
z_{i}|\balpha,\blambda,\bQ,\bX&\sim&\text{Bern}(\psi_{i})\quad\text{with}\quad
\psi_i=\Phi\left(\bx_i^\prime\balpha\right)\nonumber\\
\balpha,\blambda|\bQ,\bX&\sim&\pi, \label{eq:occbase2}
\eea
where $\Phi$ is the standard normal cdf. As it will be made evident in subsequent, we explicitly condition on $\bX$ and $\bQ$ since the priors devised for the model parameters depend on these design matrices. Again, note that the $z_i$ are not perfectly observed. The sites with $\by_i=\boldsymbol{0}$ provide no detections but this does not necessarily imply a lack of presence. Thus, the model is a zero-inflated binary regression model where both lack of presence and individual instances of detection are predicted with covariates. The observed data vectors for the sites, $\by_1,\ldots,\by_n$, are independent given $(\balpha,\blambda)$ and
\begin{align*}
p(\by_i|\balpha,\blambda,\bQ,\bX)&=\left(\Phi\left(\bx_{i}^\prime\balpha\right)\prod_j \Phi\left(\bq_{ij}^\prime\blambda\right)^{y_{ij}}\left(1-\Phi\left(\bq_{ij}^\prime\blambda\right)\right)^{1-y_{ij}}\right)^{ \mathcal{I}_{\{\by_i\neq\0\}}}
\nonumber
\\&\quad\times\left(\Phi\left(\bx_{i}^\prime\balpha\right)\prod_j \left(1-\Phi\left(\bq_{ij}^\prime\blambda\right)\right)+\left(1-\Phi\left(\bx_{i}^\prime\balpha\right)
\right)\right)^{ \mathcal{I}_{\{\by_i=\0\}}}.
\end{align*}

The model can be expanded in the spirit of \citet{ACh_93} by introducing latent variables at each level. Let $v_i$ be the underlying continuous latent variable for presence at site $i$ and $w_{ij}$ be the underlying continuous latent variable for detection during survey $j$ from site $i$. The hierarchical model in \eqref{eq:occbase2} becomes
\bea
y_{ij}|z_i,v_i,w_{ij},\balpha,\blambda,\bQ,\bX&=&z_i \mathcal{I}_{\{w_{ij}>0\}}\nonumber\\
w_{ij}|z_i,v_i,\balpha,\blambda,\bQ,\bX&\sim&N\left(\bq_{ij}^\prime \blambda,1\right)\nonumber\\
z_{i}|v_i,\balpha,\blambda,\bQ,\bX&=&\mathcal{I}_{\{v_{i}>0\}}\nonumber\\
v_i|\balpha,\blambda,\bQ,\bX&\sim&N\left(\bx_i^\prime\balpha,1\right)\nonumber\\
\balpha,\blambda|\bQ,\bX&\sim&\pi\label{eq:occbase3}.\label{gen:model}
\eea

When one uses a multivariate normal prior for $(\balpha,\blambda)$, the model in \eqref{eq:occbase3} can be fit using a Gibbs sampler. As described in \citet{DTR_12}, the only complication in using a Gibbs sampler is the fact that the sign of $v_i$ determines the value of $z_i$ and so the Gibbs sampler has to proceed in two blocks. The first block, which corresponds to a multivariate normal draw, is $(\balpha,\blambda|\bz,\bv,\bw,\by,\bQ,\bX)$. The second block is $(\bv,\bw,\bz|\balpha,\blambda,\by,\bQ,\bX)$. Each $z_i$ is drawn from the distribution $\lrsqb{z_i|\balpha,\blambda,\by,\bQ,\bX}$, which is a Bernoulli distribution with probability of success
\bean
\xi_i &=&
\mathcal{I}_{\{\by_i\neq\0\}} + \frac{\Phi\left(\bx_{i}^\prime\balpha\right)\prod_j \left(1-\Phi\left(\bq_{ij}^\prime\blambda\right)\right)}{
\Phi\left(\bx_{i}^\prime\balpha\right)\prod_j \left(1-\Phi\left(\bq_{ij}^\prime\blambda\right)\right)+1-\Phi\left(\bx_{i}^\prime\balpha\right)
} \mathcal{I}_{\{\by_i=\0\}},
\eean
and the $v_i$ and $w_{ij}$ are sampled independently from their full conditionals. Each $v_i$ has a truncated normal distribution with mean $\bx_{i}^\prime\balpha$ and variance $1$, restricted to the positive real line when $z_i=1$ and to the negative real line when $z_i=0$. Each $w_{ij}$ has a truncated normal distribution with mean $\bq_{ij}^\prime\blambda$ and variance $1$ that is supported on the positive real line when $z_iy_{ij}=1$, the negative real line when $z_i(1-y_{ij})=1$, and the whole real line when $z_i=0$. 
 
The marginal  $p(\by|\bX,\bQ)$ for the observed data can be estimated using the output from the Gibbs sampler \citep{chib1995marginal}. In this sampling scheme, one can also perform parameter expansions for both $\bv$ and $\bw$ \citep{liu1999parameter}. These dramatically decrease the autocorrelation between successive samples and reduces the asymptotic variance of estimators \citep{roy2007convergence}.
 
Alternatively, one can perform inference for the model specified in \eqref{eq:occbase2} directly using a Metropolis-Hastings algorithm (e.g., an independence chain, a random walk, or Hamiltonian Monte Carlo). The output of the Metropolis Hastings algorithm can be used to estimate the marginal of the observed data using the method outlined in \citet{chib2001marginal}. When the sample size is large, an independence chain, using the Laplace approximation to the posterior as a proposal density, provides accurate numerical estimates of the posterior evaluated at its mode in a relatively small number of samples.

\subsection{An Objective Prior for $(\balpha,\blambda)$} \label{subsec:intr-priors}

Intrinsic priors, as defined by \citet{MBR_98}, are an example of expected posterior priors \citep{Perez2002}. Concisely, an expected posterior prior for parameter $\btheta$ with prior $\pi_M$ under a model $M$ is given by
\be
\pi_M^{E}(\btheta|\pi_M,m_0)=\int p_M(\btheta|D,\pi_M)m_0(D)\d D,\nonumber
\ee
where $D$ is some imaginary data that is integrated out, $p_M(\btheta|D,\pi_M)$ is the posterior of $\btheta$ given data $D$ under the model $M$ with parameter prior $\pi_M$, and $m_0$ is a fixed distribution for generating the data $D$. The properties of the data $D$ are determined by the investigator. For regression problems, this amounts to determining the number of samples in the response and the associated design matrix. The generating model $m_0$ for the data $D$ is usually taken to be a simple model, for instance an intercept-only model. Thus, the expected posterior prior under model $M$ is calibrated to the distribution $m_0$. 

Consider the context of multiple models, $M_0,M_1,\ldots,,M_K$, where $M_0$ is nested in $M_k$ for all $k$ and model $M_k$ has parameter $\btheta_k$ with non-informative (often improper) prior $\pi^N_k$. In this context, $M_0$ is referred to as the base model. The intrinsic prior for each model is computed as
\be
\pi_{M_k}^{IP}(\btheta_k|\pi_k^N,m_0^N)=\int p^N_{M_k}(\btheta_k|D_k,\pi^N_k)m_0^N(D_k)\d D_k,
\nonumber
\ee
where $D_k$ is a training sample for model $M_k$ and $m_0^N$ is the marginal density for $D_k$ under the base model. For the intrinsic prior, $D_k$ is taken to be a minimal training sample for model $M_k$ under the prior $\pi^N_{M_k}$, which is a dataset of the smallest possible size that provides a proper posterior for $p^N_{M_k}(\btheta_k|D_k,\pi^N_k)$. Of course, the intrinsic prior for the base model is just its original non-informative prior. When the prior for model $M_k$ is improper and only defined up to a multiplicative constant $c_k$, the intrinsic prior framework removes the ambiguity of these constants and each intrinsic prior is defined up to a common multiplicative constant $c_0$.

An extension of the intrinsic prior framework is to have the datasets $D_k$ include both observable and unobservable latent variables. \citet{Leon-Novelo2012} used this approach in computing an objective prior for standard probit regression. There, the authors conditioned on both the observed binary data as well as the unobserved continuous latent variables. Following their development, we form an objective prior for the occupancy model by conditioning on the unobserved latent presence variables ($\bz$) as well as the unobserved continuous latent variables for both presence and detection ($\bv,\bw$). We refer to this objective prior as an intrinsic, prior though its derivation differs from that in \citet{MBR_98} and \citet{Berger1996}.

Specifically, let $\bX_0$ and $\bQ_0$ be design matrices for presence and detection in the model $M_0$ and let $\bX$ and $\bQ$ be design matrices for a model $M$ that nests $M_0$. Let $(\balpha,\blambda)$ and $(\balpha_0,\blambda_0)$ be the parameters of $M$ and $M_0$, respectively. Further, assume that the prior distributions for the parameters under each model are constant, $\pi^N_M=c_M$ and $\pi^N_0=c_0$. The intrinsic prior for $(\balpha,\blambda)$ is given by
\be
\pi^{IP}_M(\balpha,\blambda|\tilde{\bQ},\tilde{\bX})=\sum_{\tilde{\bz},\tilde{\by}}\iint 
p^N_M(\balpha,\blambda|\tilde{\bz},\tilde{\bv},\tilde{\bw},\tilde{\by},\tilde{\bQ},\tilde{\bX}) m_0^N(\tilde{\bz},\tilde{\bv},\tilde{\bw},\tilde{\by}|\tilde{\bQ}_0,\tilde{\bX}_0)
\d \tilde{\bv}\,\d\tilde{\bw},\label{eq:ip1}
\ee
where the ``$\sim$'' over variables indicates that these correspond to the training sample that is to be integrated out. The formula in \eqref{eq:ip1} is greatly simplified by the fact that, under the non-informative prior, $(\balpha,\blambda)$ are conditionally independent of $(\tilde{\bz},\tilde{\by})$ given the continuous latents $(\tilde{\bv},\tilde{\bw})$. Moreover, the $\balpha$ and $\blambda$ are conditionally independent of each other given the continuous latents. Thus, \eqref{eq:ip1} simplifies to
\begin{align}
\pi^{IP}_M(\balpha,\blambda|\tilde{\bQ},\tilde{\bX})
&
=\iint 
p^N_M(\balpha|\tilde{\bv},\tilde{\bw},\tilde{\bQ},\tilde{\bX})
p^N_M(\blambda|\tilde{\bv},\tilde{\bw},\tilde{\bQ},\tilde{\bX}) m_0^N(\tilde{\bv},\tilde{\bw}|\tilde{\bQ}_0,\tilde{\bX}_0)
\d \tilde{\bv}\,\d\tilde{\bw}\nonumber\\
&
=\int 
p^N_M(\balpha|\tilde{\bv},\tilde{\bX})m_0^N(\tilde{\bv}|\tilde{\bX}_0)\d \tilde{\bv}\times
\int p^N_M(\blambda|\tilde{\bw},\tilde{\bQ}) m_0^N(\tilde{\bw}|\tilde{\bQ}_0)
\d\tilde{\bw},\label{eq:ip2}
\end{align}
where the last equality follows from the assumptions of \eqref{eq:occbase3} and the prior independence of $\balpha$ and $\blambda$ under $\pi^N_M$. 
Both of the integrals in \eqref{eq:ip2} are of the form of the integrals in \citet{Leon-Novelo2012}. Thus, the intrinsic prior is given by a product of singular normal distributions. 

The explication of these priors is greatly aided by the introduction of additional notation. Because $M_0$ is nested in $M$, we can write $\bX=\lrp{\bX_0\quad \bX_A}$ and $\bQ=\lrp{\bQ_0\quad \bQ_A}$ and can do the same for the design matrices for the minimal training sample. Similarly, we can write $\balpha=(\balpha_0^\prime, \balpha_A^\prime)^\prime$ and $\blambda=(\blambda_0^\prime, \blambda_A^\prime)^\prime$. The intrinsic prior is given by
\bea
\balpha_A|\balpha_0,\tilde{\bX} &\sim& \mathcal{N}\left(\0, \,2\lrp{\tilde{\bX}_A^\prime\lrp{\bI-\tilde{\bH}_{0_z}}\tilde{\bX}_A}^{-1}\right)\label{eq:womprior1}\\
\blambda_A|\blambda_0,\tilde{\bQ} &\sim& \mathcal{N}\left(\0,\,2\lrp{\tilde{\bQ}_A^\prime\lrp{\bI-\tilde{\bH}_{0_y}}\tilde{\bQ}_A}^{-1}\right)\label{eq:womprior2}\\
\blambda_0,\balpha_0|\tilde{\bX},\tilde{\bQ} &\sim& c_0 \times d_0 \label{eq:womprior3}
\eea
where $\tilde{\bH}_{0_z}$ and $\tilde{\bH}_{0_y}$ are the hat matrices associated to $\tilde{\bX}_0$ and $\tilde{\bQ}_0$, respectively. Here we include two constants $c_0$ and $d_0$ for the reference prior for the base model, where $c_0$ and $d_0$ are undefined constants for the flat priors for $\balpha_0$ and $\blambda_0$, respectively.

The only remaining task for this intrinsic prior is to define the design matrices for the minimal training samples. Letting $p_\balpha=\dim(\balpha)$ and $p_\blambda=\dim(\blambda)$, the minimal training samples for $\bv$ and $\bw$ contain $p_\balpha$ and $p_\blambda$ samples, respectively. Following \citet{Leon-Novelo2012} and \citet{Casella2006}, we define $\tilde{\bX}$ and $\tilde{\bQ}$ to be any design matrices of dimensions $p_\balpha\times p_\balpha$ and $p_\blambda\times p_\blambda$ satisfying
\be
\tilde{\bX}^\prime\tilde{\bX}=\frac{p_\balpha}{N} \bX^\prime\bX \qquad\text{and}\qquad \tilde{\bQ}^\prime\tilde{\bQ}=\frac{p_\blambda}{\jd} \bQ^\prime\bQ,
\ee
where $N$ is the number of sites and $\jd=\sum_{i=1}^{N}J_{i}$ is the total number of surveys.
Note that the covariance matrices in \eqref{eq:womprior1} and \eqref{eq:womprior2} are thus completely determined by $\bX^\prime\bX$ and $\bQ^\prime\bQ$.

\section{The Variable Selection Problem}\label{subsec:notation}

The hierarchy in Equation \eqref{gen:model} is given for a specific model with a fixed set of predictors. 
This section develops the model selection problem for occupancy models. Each model contains two components, one for presence and one for detection. Thus, model $M$ is decomposed as $M=\lrp{M_y,M_z}$, where $M_y$ is a component model for detection and $M_z$ is a component model for presence. The base model is $M_0=\lrp{M_{0_y}, M_{0_z}}$, where the component base model design matrices contain at least a column of ones for the intercept. Each model $M$ is assumed to nest $M_0$ and the prior for model $M$ is taken to be the intrinsic prior defined in \eqref{eq:womprior1}-\eqref{eq:womprior3}. The largest model is denoted by $\MF=\lrp{\MFsub{y},\MFsub{z}}$ and contains the largest possible component models for detection and presence. The design matrices for these full components are $\bX_F$ and $\bQ_F$.

Let $K=(K_y, K_z)$, where $K_y$ and $K_z$ denote the sets of column indices for $\bQ_{F}$ and $\bX_{F}$ that are not in $\bQ_0$ and $\bX_0$, respectively. The model space can then be represented by the Cartesian product $\mathcal{P}\lrp{K_y}\times \mathcal{P}\lrp{K_z}$, where $\mathcal{P}\lrp{B}$ is the powerset of $B$. A specific model is represented by $A=\lrp{A_y, A_z}$ with $A_y\subseteq K_y$ and $A_z\subseteq K_z$. Thus, the entire model space $\mathcal{M}$ is populated by models of the form $M_{A}=\lrp{M_{A_y},M_{A_z}}$, where $M_{A_y}$ and $M_{A_z}$ are the corresponding component models for detection and presence determined by the base covariates as well as covariates with indices in $A_y$ and $A_z$, respectively. It follows that for the presence process $\bz$, the design matrix for the model $M_{A}$ is of the form $\bX_{M_A}=\lrp{\bX_{0}\quad \bX_{A}}$, where $\bX_{0}$ is the design matrix of the base model $M_{0_z}$ and $\bX_{A}$ is the matrix containing the covariates indexed by $A_z$ (and similarly for $\bQ_{M_A}=\lrp{\bQ_{0}\quad \bQ_{A}}$). Denote the regression coefficients of the model $M_A$ by $\balpha_{M_A}=(\balpha_0^\prime,\balpha_A^\prime)^\prime$ and $\blambda_{M_A}=(\blambda_0^\prime,\blambda_A^\prime)^\prime$ for presence and detection, respectively.

It is important to note that this construction using the Cartesian product provides the largest possible model space for the occupancy model given the structures of the base and full models. Investigators may wish to impose additional model space restrictions based upon their (subjective) judgment. One means of achieving this restriction is to form two sets of models, $\mathcal{M}_y$ for detection and $\mathcal{M}_z$ for presence. The model space can then be defined by the Cartesian product, $\mathcal{M}=\mathcal{M}_y\times\mathcal{M}_z$. One particular example of such a restriction arises when higher-order terms are included in the detection or presence models. Heredity conditions \citep{Chipman1996} can be imposed on either model space and appropriate priors defined (see Section \ref{subsec:model-priors}).

\subsection{Priors over the Space of Models}\label{subsec:model-priors}
Here we outline the construction of prior distributions over the model space. To allow  for flexible modeling, it is assumed that the sets of covariates can potentially include interaction effects, higher-order polynomial terms, and factor variables. The priors for either the presence or the detection component have the same structure, and the joint prior is the product of marginal priors of the two model components. 

The priors placed on the model space for the presence and detection models respect the hierarchy of the terms that could be included in a given model. Aspects of the prior construction are described here and full details on such priors can be found in \citet{Taylor-Rodriguez2015}. The full model for either the presence or the detection component is represented as a directed acyclic graph (DAG) with nodes representing polynomial terms (powers or interactions; e.g., $x_1$ or $x_1^2$ or $x_1 x_2^2$) and with edges specifying inheritance relationships.  For example, $x_1 x_2^2$ has edges (inherits) from its parent nodes $x_1 x_2$ and $x_2^2$, also $x_1^2$ inherits from its parent $x_1$ but not from $x_2$. Feasible models, also known as models obeying weak heredity, correspond to a special kind of connected subgraph of the full model DAG. First, they must include the base model DAG. Second, a node $\eta$ can only be included in a model's subgraph only if there is a directed path from a node in the base model to $\eta$. The priors considered here focus on models satisfying a strong heredity (also known as well-formulated models), which amounts to requiring that for each node $\eta$ in a model's subgraph, all parents of $\eta$ included in the model's subgraph. 

Model prior probabilities are specified recursively via conditional node inclusion probabilities (given the parent DAG) using a type of Markov condition reflected in the principles of conditional independence and immediate inheritance \citep{Chipman1996}. Conditional node inclusion is identified with a latent Bernoulli random variable and placing a conjugate (beta) prior on the inclusion probabilities. The model space prior is obtained by integrating out the conditional inclusion probabilities. In the simplest case all of conditional inclusion probabilities are assumed to be equal and the prior is called the hierarchical uniform prior (HUP). The amount of penalization of complex models can be adjusted (typically, increased relative to the purely combinatorial penalization of the HUP) using node-specific inclusion probabilities and stronger shrinkage via the beta hyper-priors on the inclusion probabilities; this results in the hierarchical independence (HIP) and order priors (HOP) that group nodes of similar complexity together. % For full details, see \citet{Taylor-Rodriguez2015}. 

\subsection{Model Posterior Probabilities}\label{subsec:post-prob}

In order to compute the posterior probabilities of interest, we take advantage of the model representation making use of the latent variables introduced for the presence and detection processes.
%, and work the model selection problem in the latent normal scale. 
Specifically,
%following a  strategy similar to that of \citet{Womack2012} for ordinal probit regression, 
a conditional independence argument provides
\begin{eqnarray}
p(M_{A}|\by,\bz,\bw,\bv)&=&\frac{m(\by,\bz,\bw,\bv|M_A)\pi(M_A)}{m(\by,\bz,\bw,\bv)}\nonumber\\
&=& \frac{f_{\by,\bz}(\by,\bz|\bw,\bv) m(\bw,\bv|M_{A}) \, \pi(M_A)} {f_{\by,\bz}(\by,\bz|\bw,\bv) \sum_{M^*\in\cM}  m(\bw,\bv|M^*)\pi(M^*)}\nonumber\\
&=&\frac{m(\bw,\bv|M_{A})\pi(M_A)}{m(\bw,\bv)},
\label{eq:PostWLat}
\end{eqnarray}
because $\bz$ is independent of $M_A$ once $\bv$ is known and $\by$ is independent of $M_A$ once $\bz$ and $\bw$ are known. In \eqref{eq:PostWLat}, 
\bea
 f_{\by,\bz}(\by,\bz|\bw,\bv)&=&\prod_{i=1}^{N}\mathcal{I}_{\{v_{i}>0\}}^{z_{i}}\mathcal{I}_{\{v_{i}\leq0\}}^{(1-z_{i})} \prod_{j=1}^{J}(z_{i}\mathcal{I}_{\lrb{w_{ij}>0}})^{y_{ij}}(1-z_{i}\mathcal{I}_{\lrb{w_{ij}>0}})^{1-y_{ij}},\nonumber\\
m(\bw,\bv|M_A)&=& m(\bv|M_{A_z}) m(\bw|M_{A_y})\nonumber\\
&=&\int \int \lrp{\prod_{i=1}^{N}\phi(v_{i}|\bx_{i}^{\prime}\balpha,1; M_{A_{z}})}  \lrp{\prod_{i=1}^{N}\prod_{j=1}^{J_i}\phi(w_{ij}|\bq_{ij}^{\prime}\blambda,1; M_{A_{y}})} \times\nonumber\\
&&\nonumber\\
&&\qquad\pi^{IP}_{M_A}(\balpha,\blambda|\tilde{\bQ},\tilde{\bX}) \d\balpha \d\blambda,
\label{eq:PostOddsLat2}
\eea
with $\phi(\cdot|\bmu,\sigma^2; M)$ denoting the normal pdf with mean $\bmu$, variance $\sigma^2$ conditional on model $M$, and  $\pi^{IP}_{M_A}(\balpha,\blambda|\tilde{\bQ},\tilde{\bX})$ as defined in \eqref{eq:ip2}.
% Closed-form expressions for the marginals $m(\bw|M_{A_y})$ and $m(\bv|M_{A_z})$ are derived in as follows: 

Under the intrinsic priors above, the closed-form expression for the marginal $m(\bv|M_{A_z})$ is
 \begin{eqnarray}
 m(\bv|M_A)% &=& \int \int c_0\,\mathcal{N}\lrp{\bv|X_0\balpha_0+X_{A}\balpha_{A},\mathcal{I}} \mathcal{N}\lrp{\balpha_{A}|\0,2(\tilde{X}_{A}^\prime\tilde{X}_{A})^{-1}}d\balpha_{A} d\balpha_0 \nonumber\\
 %&=& c_0(2\pi)^{-n/2}\,\int \lrp{\frac{p_{A_z}}{2N+p_{A_z}}}^{\frac{(p_{A_z}-p_{0_z})}{2}} \times{}\nonumber\\
% &&{}\exp{-\frac{1}{2}\lrp{\bv-X_{0}\balpha_0}^\prime\lrp{\mathcal{I}-\lrp{\frac{2N}{2N+p_{A_z}}}H_{A}^{(z)} } \lrp{\bv-X_{0}\balpha_0} }\d\balpha_0\nonumber\\
 &=&c_0\,(2\pi)^{-(n-p_{0_z})/2} \lrp{\frac{p_{A_z}}{2N+p_{A_z}}}^{\frac{(p_{A_z}-p_{0_z})}{2}} \left|\bX_0^\prime \bX_0 \right|^{-\frac{1}{2}} \times{}\nonumber\\
 &&{} \exp{-\frac{1}{2}\bv^\prime\lrp{\bI - \bH_{0_z}-\lrp{\frac{2N}{2N+p_{A_z}}}\bH_{A_z}^\perp }\bv},\label{eq:margbv}
 \end{eqnarray}
 where $\bH_{A_z}^{\perp}$is the hat matrix associated with $(\bI-\bH_{0_z})\bX_A$.   Similarly, the marginal distribution for $\bw$ under model $M_A$ is
 \begin{eqnarray}
 m(\bw|M_A) &=& d_0\,(2\pi)^{-(\jd-p_{0_y})/2} \lrp{\frac{p_{A_y}}{2\jd+p_{A_y}}}^{\frac{(p_{A_y}-p_{0_y})}{2}} \left|\bQ_0^\prime \bQ_0 \right|^{-\frac{1}{2}} \times{}\nonumber\\
 &&{} \exp{-\frac{1}{2}\bw^\prime\lrp{\bI-\bH_{0_y}-\lrp{\frac{2\jd}{2\jd+p_{A_y}}}\bH_{A_y}^{\perp} }\bw},\label{eq:margbw}
 \end{eqnarray}
where $\bH_{A_y}^{\perp}$ is the hat matrix associated with $(\bI-\bH_{0_y})\bQ_{A}$ and $\jd=\sum_{i=1}^{N}J_{i}$ is the total number of surveys.  Finally, the marginals for the base model $\M_0=\lrp{M_{0_y},M_{0_z}}$ are
% \begin{eqnarray}
% m(\bv|M_0) &=& \int c_0\,\mathcal{N}\lrp{\bv|X_0\balpha_0,\bI} \d\balpha_0 \nonumber\\
% &=& c_0(2\pi)^{-(n-p_{0_z})/2}\left|\bX_0^\prime \bX_0 \right|^{-\frac{1}{2}}\exp{-\frac{1}{2}\lrp{\bv'\lrp{\bI-\bH_{0_z} }\bv}}\label{eq:MargM0v}
% \end{eqnarray}
% and
% \begin{eqnarray}
% m(\bw|M_0) &=& d_0(2\pi)^{-(\jd-p_{0_y})/2}\left|\bQ_0^\prime \bQ_0 \right|^{-\frac{1}{2}}\exp{-\frac{1}{2}\lrp{\bw'\lrp{\bI-\bH_{0_y} }\bw}}.\label{eq:MargM0w}
% \end{eqnarray}
 
\begin{eqnarray}
 m(\bv|M_0) &=& \int c_0\,\mathcal{N}\lrp{\bv|X_0\balpha_0,\bI} \d\balpha_0 \nonumber\\
 &=& c_0(2\pi)^{-\frac{(n-p_{0_z})}{2}}\left|\bX_0^\prime \bX_0 \right|^{-\frac{1}{2}}\exp{-\frac{1}{2}\lrp{\bv'\lrp{\bI-\bH_{0_z} }\bv}}\label{eq:MargM0v}\\
 \mbox{and}  \nonumber \\
 m(\bw|M_0) &=& d_0(2\pi)^{-\frac{(\jd-p_{0_y})}{2}}\left|\bQ_0^\prime \bQ_0 \right|^{-\frac{1}{2}}\exp{-\frac{1}{2}\lrp{\bw'\lrp{\bI-\bH_{0_y}}\bw}}.\label{eq:MargM0w}
 \end{eqnarray}
The specification of the model posteriors in Equation \eqref{eq:PostWLat} is completed using the construction of the priors $\pi(M_A)$ over the model space; see Section \ref{subsec:model-priors}.

The advantage of \eqref{eq:PostWLat} is that the posterior of model $M_A$ can be represented as 
\begin{equation}
p(M_A|\by)=\iiint p(M_A|\by,\bz,\bw,\bv)f(\bz,\bw,\bv|\by)\d\bz\d\bw\d\bv, \label{eq:postint}
\end{equation}
which provides for straightforward ergodic estimation of $p(M_A|\by)$ if samples can be drawn from $f(\bz,\bw,\bv|\by)$. If $S$ such samples are obtained, then \eqref{eq:postint} can be approximated by 
\begin{equation}
S^{-1}\sum_\ell p(M_A|\by,\bz^{(\ell)},\bw^{(\ell)},\bv^{(\ell)}).\label{eq:postest}
\end{equation}
Such draws can be obtained using reversible jump MCMC (RJMCMC) \citep{Green1995}, as described in the Appendix. One subtle point of difficulty is the calculation of $m(\bw,\bv)=\sum_{M_A} m(\bw,\bv|M_A)\pi(M_A)$ in the denominator of \eqref{eq:PostWLat} when the space of models is too large to be enumerated (or if the necessary calculations for each model and each draw of $(\bw,\bv)$ are too arduous). In such a case, the sum may be approximated by $T^{-1}\sum_t m(\bw,\bv|M^{(t)})\pi(M^{(t)})$, where $t$ indexes a set of $T$ models. For instance, $t$ could index the set of models visited during the RJMCMC sampler or a larger set of models could be used (the posterior of a model $M_A$ not in this set can be estimated using \eqref{eq:postest}).

% \subsection{Model Posterior Probabilities}\label{subsec:post-prob}

%This result implies that once the occupancy and detection indicators are conditioned on the latent processes $\bv$ and $\bw$, respectively, the model posterior probabilities are proportional to the marginals for the latent variables conditional on the model. Hence, the model selection problem is driven by the posterior probabilities
%\begin{eqnarray}
%p(M_{A}|\by,\bz,\bv,\bw)&\propto& m(\bw|M_{A_y})\,  m(\bv|M_{A_z})\, \pi(M_A). \label{eq:PostOddsLat}
%\end{eqnarray}
%% \vspace{0.2cm}
%\red{NB: this is slightly misleading since the selection is also driven by the marginal posterior $f(\bz,\bv,\bw|\by)$.} \textcolor{blue}{(how would you suggest we clarify this?)}\\

%From expression \eqref{eq:PostWLat}, recall that 
%$$p(M_{A}|\by,\bz,\bv,\bw)\propto m(\bw|M_{A_y})\,  m(\bv|M_{A_z})\, \pi(M_A),$$
%and that estimation of $p(M_A|\by)$ can be achieved using draws of $(\bz,\bv,\bw)$ from the RJMCMC sampler described in the Appendix.

%where only $\bz_o$ are observed, the model selection is driven by $p(M_{A}|\by)=\text{E}_{\bw,\bv,\bz|\by}\big\{ p(M_{A}|\by,\bz,\bw,\bv)\big\}$. In practice, we estimate these posterior probabilities by drawing $(\bw,\bv,\bz)\sim p(\bw,\bv,\bz|\by)$ (see appendix A.1) and obtaining the Monte Carlo average of $m(\bw,\bv|M_A)$. Given that $\bw$ and $\bv$ are independent given $M_A$, in order to obtain the model posteriors from (\ref{eq:PostWLat}) we derive closed-form expressions for the marginals $m(\bv|M_{A_z})$ and $m(\bw|M_{A_y})$.  

\section{Simulation Experiments}\label{sec:sims}

This section considers nine different scenarios where we explore a range of detectability and prevalence regimes to assess the behavior of the proposed algorithm.  For each model component, the base model is taken to be the intercept-only model, and the full models considered for the presence and the detection have, respectively, five and three predictors.  Therefore, the model space contains $2^{5}\times 2^{3}=256$ candidate models.  The assumed true models are $\M_{Tz}=\lrb{1, x_{1},x_{2} ,x_{5}}$ for the presence  and $\M_{Ty}=\lrb{1, q_{2},q_{3}}$ for the detection, where $1$ represents the intercept. This small model space is considered so that comparisons with selection using AIC (which generally requires complete enumeration of the model space) can be made.

The simulation scenarios we consider vary depending on where the distributions for the detection and presence probabilities are centered. That is, we set the average probability for detection and presence to predefined values $\bar{p}$ and $\bar{\psi}$, respectively.   If the detection probabilities are centered near one, a non-detection commonly implies a non-presence since the detection is almost perfect.  On the contrary, if the detection probabilities are centered close to zero (as with cryptic species), then the uncertainty surrounding an observed zero is greater, making it more difficult to determine if this also corresponds to a true zero in the presence.  Now, combining the different values for $\bar{p}$ with different values for the center of the distribution for the presence probabilities $\bar{\psi}$, we can account for a variety of possibilities observed in real data, ranging from cryptic but highly prevalent species, to easy to detect but very rare species.

The mean probability values for detection and presence that determine our scenarios correspond to the pairs $(\bar{p}, \bar{\psi})\in\{0.2,0.5,0.8\}^{\times 2}$.  To match the target values $(\bar{p}, \bar{\psi})$, 15 independent sets of $\lrb{\bX_F,\bQ_F}$ were drawn independently from the standard normal distribution, and for each of them the true model parameters were chosen to solve for $\balpha$ and $\blambda$ the equations $\hat{\psi}(\balpha)=\bar{\psi}$ and $ \hat{p}(\blambda)=\bar{p}$, where 
\bean
\hat{\psi}(\balpha)&=&\frac{1}{N}\sum_{i=1}^{N} \Phi(\bx_i^\prime \balpha)\text{ and} \\
\hat{p}(\blambda)&=&\frac{1}{\sum_{i=1}^{N} J_i}\sum_{i=1}^{N}\sum_{j=1}^{J_i} \Phi(\bq_{ij}^\prime \blambda).
\eean
For each scenario and dataset combination, we used the best solution from ten runs of a gradient-based (quasi-Newton) algorithm initialized from independent standard normal draws.  Finally, having determined the regression parameters corresponding to the different scenarios and conditioning on $M_{Tz}$ and $M_{Ty}$, the true presence and detection indicators were drawn from the probit model described by \eqref{eq:occbase2} for each dataset.

\begin{figure}[!h]
\begin{center}
\centering
\subfigure[Presence]{\includegraphics[scale=.55]{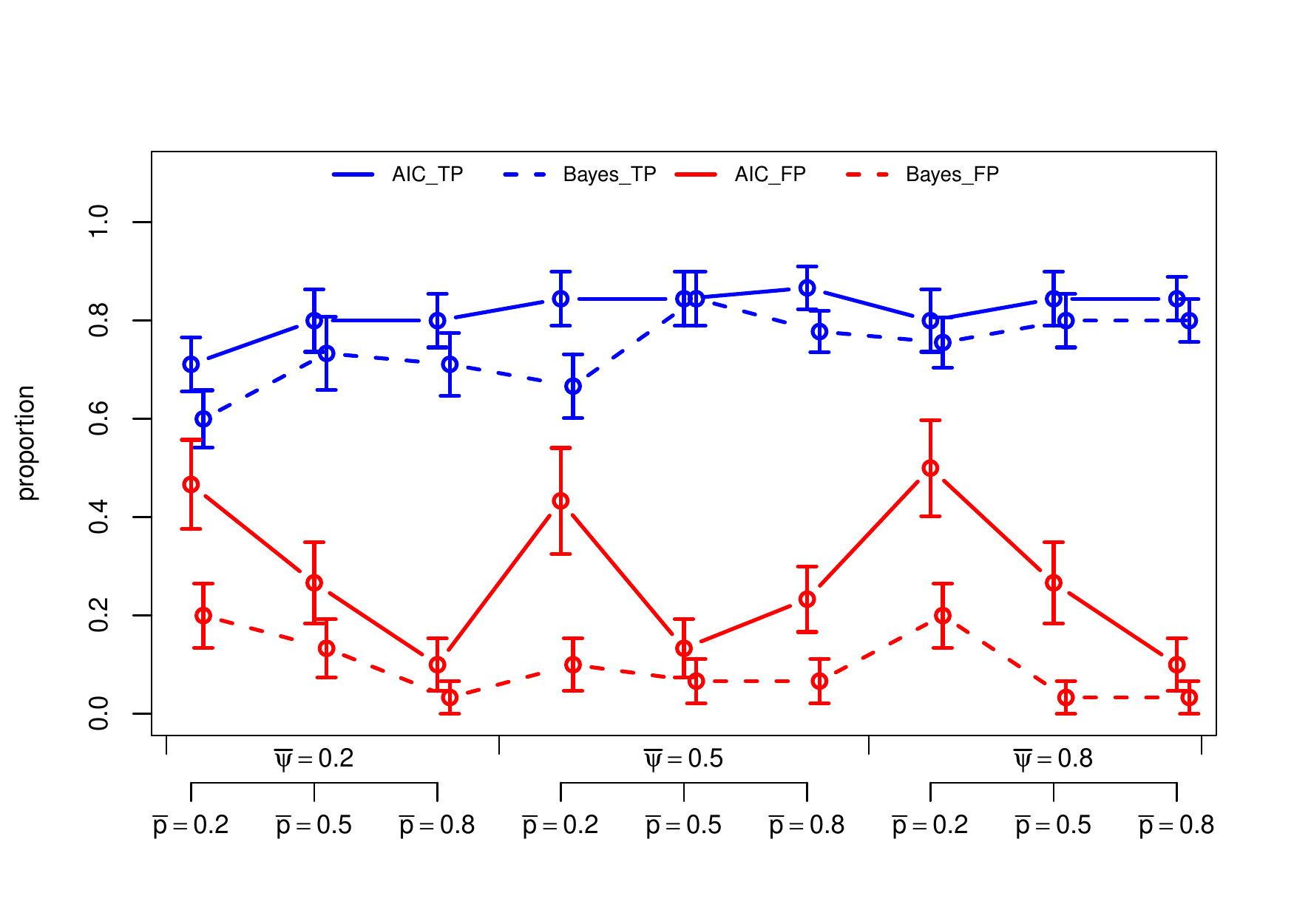}}\\
\subfigure[Detection]{\includegraphics[scale=.55]{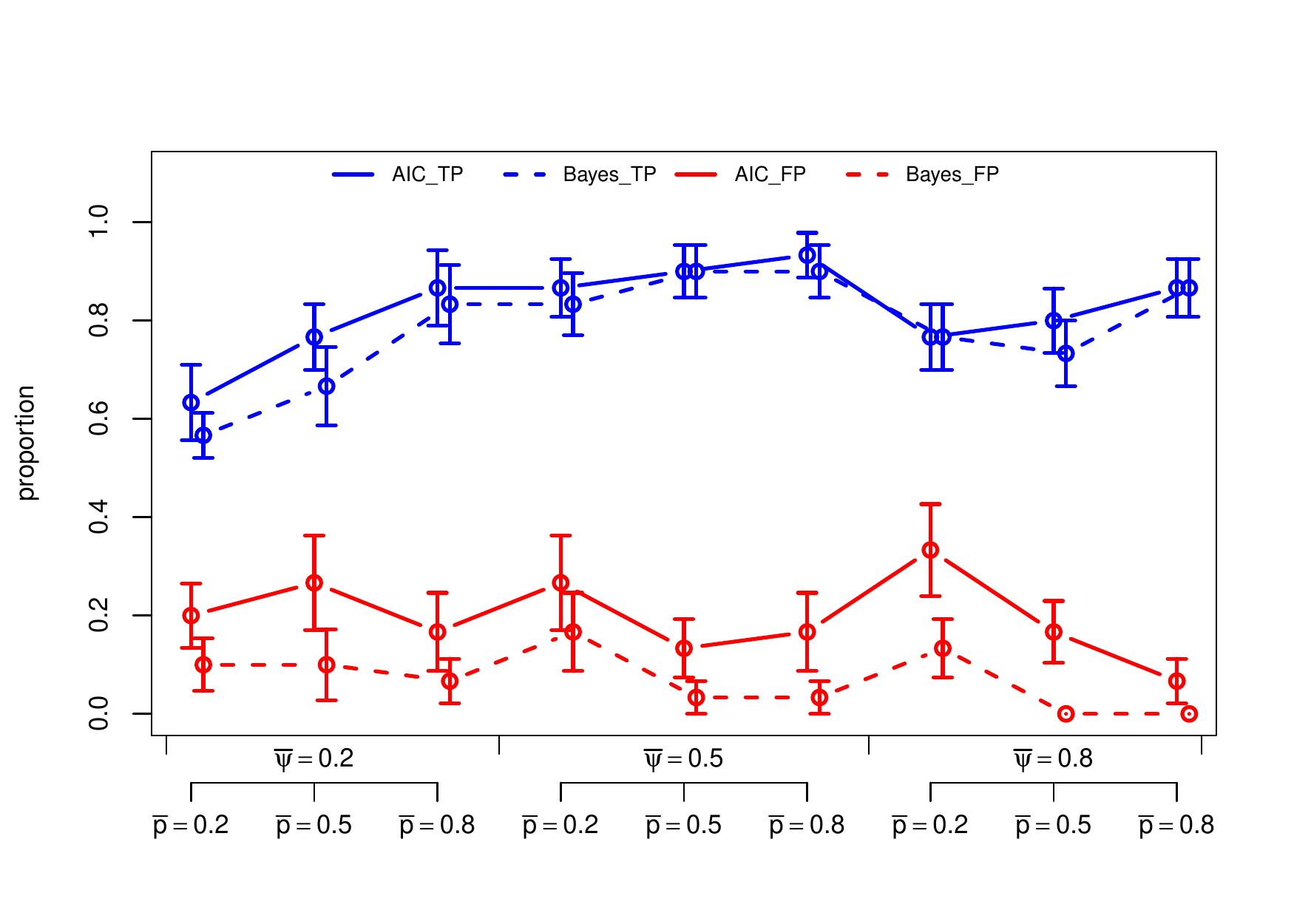}}
\end{center}
\caption{Proportion of true positives (TP) and false positives (FP) using the proposed approach and AIC for the detection and the presence components of the model.}
\label{fig:TPFP}
\end{figure}

% ---------- mod -----

The results are shown in Figure \ref{fig:TPFP},  which depicts the average proportion of true positive (TP) and false positive (FP) predictors included in the selected models under each scenario.  The TP predictors are those in the true model that are also in the selected model, and the FP predictors correspond to those absent from the true model but included in the chosen model.  The selected models are the lowest AIC model and the median probability model (MPM) under the objective Bayes methodology.  The MPM is the model that includes all predictors whose marginal posterior inclusion probability (MPIP) is greater than or equal to 0.5, where the MPIP for a given predictor is defined as
\bea
p(\text{predictor is included} | \by) &=&\sum_{M\in\cM}p(M|\by,\cM)\mathcal{I}_{\lrb{\text{predictor}\in\M}}.\label{eq:MPIPs}
\eea
The TP and FP rates for both detection and presence components lead to the same conclusions.  In terms of the TPs, the AIC selects a slightly higher number of true positive terms, especially for the component of the model associated to the presence indicators.  Nonetheless, these differences are modest at most.  Conversely, the resulting proportions of false positive terms (FP) tend to be strinkingly lower using our method, especially for the presence component in those scenarios where there is poor detection (i.e., $\bar{p}=0.2$).  Remarkably, whenever the species is highly prevalent ($\bar{\psi}=0.8$) and detection ranges between moderate and high ($\bar{p}=0.5,0.8$), the number of false positive terms under our approach is very close to zero in both model components. Also, with $(\bar{p},\bar{\psi})=(0.2, 0.8)$ our method substantially outperforms AIC in filtering out the false positive terms both in the presence and detection components.

These results are very encouraging: the proposed method not only reduces the inclusion of false positive terms in comparison to AIC but also has comparable performance finding true predictors.  
%%%%%%%%%%%%%%%%%%%%%%%%%%%%%%%%%%%%%%%%%%%%%%%%%%%%%%%%%%%%%%

\section{Case studies}\label{sec:applications}

In this section, we analyze two datasets. First, we consider presence-absence data for mallard wild ducks ({\sl Anas platyrhynchos}), collected as part of the 2002 Swiss breeding bird monitoring program. For our second example, we consider the blue hawker dragonfly data, which had been previously studied using AIC as the variable selection strategy in \citet{KG_10}.  The mallard data is extremely clean, with sufficient sites being surveyed, which for the most part are visited the same number of times.  On the other hand, the blue hawker dataset was collected through a large scale citizen science effort.  As such, although the number of sites visited is large for this type of data, it displays large asymmetries in the surveying effort, posing a more challenging problem for this type of analysis. 

Both data sets contain a sufficiently small number of predictors so that enumeration of the entire model space is feasible. Therefore, for these data analyses, we present estimators of posterior probabilities from enumeration (EPE), renormalization (RPE), and visit frequency (FPE). While all estimates exhibit Monte Carlo error, we treat the enumeration estimators as a gold standard estimator because the Monte Carlo error can be easily controlled. We implement the method of \citet{chib2001marginal} for estimation of the marginal and use a relative magnitude stopping rule to determine the length of sampling \citep{flegal2013relative}. In particular, we require that the $95\%$ confidence interval for the estimator for the log posterior evaluated at its mode be less than $1\%$ of the size of the estimate.

To obtain the EPE for each model, we run the MCMC algorithm defined by  \eqref{gen:model} using the priors given in \eqref{eq:womprior1}-\eqref{eq:womprior3}.  These yield draws of the regression coefficients conditional on each model, which are then used to calculate the marginal density of the response.  We calculate the EPEs using the marginals obtained under each model.  Once the EPEs are in place, we then compare them to their corresponding MCMC estimates using either FPE or RPE.  Expression \eqref{eq:postest} enables direct calculation of the RPEs for a specified set $\cM_A$ of models, which may even include models that were not sampled. Given the moderate size of the model space for these examples, in both cases we set $\cM_A$ to be the entire model space. In contrast, as a general rule the FPEs are only available for the set of the visited models in the RJMCMC.  Finally, to compare our results to the traditional approach using AIC, we use the ``Akaike weights" \citep[see][for a definition and further information]{burnham2003model,Burnham2004}. These are obtained using functions \textsf{occu} and \textsf{dredge} from the R packages \textsf{unmarked} and \textsf{MuMin}, respectively.  The AIC weights allow us to make direct comparison of the results provided by either method, as they can be seen as posterior probabilities obtained from a specific prior on the model parameters.  However, as AIC is minimax-rate optimal for estimating the regression function, it cannot be a consistent model selector, as demonstrated in \citet{Yang2005}, making these priors ill-suited for variable selection.

\subsection{The mallard data}\label{subsec:mallard}

As Switzerland is a small and mountainous country, it provides for large variation in its topography and physio-geography. As such, elevation is a good candidate to predict species occurrence at a large spatial scale.  It can serve as a proxy for habitat type, intensity of land use, temperature, as well as some other biotic factors \citep{KG_10}.   The data used in the illustration was collected by the Swiss breeding bird survey, and had been previously used to derive abundance estimates in \citet{Kery2005}.  

 The monitoring program for common breeding bird species comprises more than 250 1-km$^2$ quadrats distributed in a grid sample across Switzerland.  Throughout the breeding season, each quadrat is surveyed two or three times annually by an experienced surveyor along a route, recording the date and whether visual or acoustic contact was made. Elevation (elev) and forest cover (forest) were matched for the studied locations from the Swiss Federal Statistical Office \citep{Kery2004}.  Given that the route length (length) across quadrats was not homogenous, route length (within a quadrat) was considered to account for variation in effective sample area.  To model the detection probabilities, survey duration  divided by route length (ivel) was used as a measure of effort.  Also the date (date) was considered for the detection component since the surveys were collected over a three month period, and behavioral changes that might affect detection could be expected.  Using the built-in feature of our algorithm to account for the polynomial structure in the predictors, we considered a full quadratic surface for the predictors, both in the presence as well as in the detection component.  The dataset contains 235 quadrats, of which two were surveyed once, 42 twice, and 191 were visited three times.

\subsubsection{Results}

As mentioned above, given that this dataset contains only a few covariates, even when considering the full quadratic surfaces, it is possible to perform complete enumeration of the model space (which has 1,235 models).  The results from our analyses are summarized in terms of the MPIPs (calculated using \eqref{eq:MPIPs}), the top ranked models (in terms of their posterior probabilities), and the Median Probability Model (MPM), which is the model containing only terms whose MPIPs are greater than 0.5.  These measures were all obtained for each method using the posterior probabilities from the joint model for presence and detection.  

Table \ref{tab:MPIPsMallard} displays the MPIPs calculated with EPEs, RPEs, FPEs and AIC$_w$. Although the MPIPs obtained from EPE are lower than those from the two other estimates (RPE and FPE), for the most part all three share the same ordering, with the exception of the \textsf{length}$^2$ term in the presence component. It is worth noting that, although the MPIPs are comparable for the three alternatives, those from RPE are considerably closer to those from EPE than those from RPE, especially for the detection component. The MPIPs from AIC$_w$ are considerably higher for most predictors than any of their Bayesian counterparts, implying that good models resulting from AIC selection are more complex, as expected.
\begin{table}[!ht]
\caption{MPIPs from joint model for the presence (top) and the detection (bottom) components for the mallard dataset}\label{tab:MPIPsMallard}
\centering
\begin{tabular}{lcccc}
\toprule
 & EPE & RPE & FPE & AIC$_w$\\ 
\midrule
\textsf{elev} & 0.9966 & 1.0000 & 1.0000 & 1.0000 \\ 
\textsf{forest} & 0.9446 & 0.9525 & 0.9489 & 0.9987\\ 
\textsf{length} & 0.4305 & 0.5998 & 0.5983 & 0.9625 \\ 
\textsf{length*forest} & 0.2153 & 0.3803 & 0.4090 & 0.8737\\ 
\textsf{elev*length} & 0.2069 & 0.3336 & 0.3491 & 0.7561\\ 
\textsf{elev*forest} & 0.1297 & 0.1448 & 0.1732 & 0.3577 \\ 
\textsf{elev$^2$} & 0.1110 & 0.1293 & 0.1620  & 0.3347\\ 
\textsf{forest$^2$} & 0.1067 & 0.1229 & 0.1504 & 0.3077\\ 
\textsf{length$^2$} & 0.0734 & 0.1440 & 0.1639 & 0.5333\\ 
\bottomrule
 &  &  &  & \\ 
\toprule
 & EPE & RPE & FPE & AIC$_w$\\ 
\midrule
\textsf{date} & 0.1315 & 0.1982 & 0.3846 & 0.5573\\ 
\textsf{ivel} & 0.0538 & 0.1476 & 0.3568 & 0.3086\\ 
\textsf{date$^2$} & 0.0258 & 0.0560 & 0.1119 & 0.3645\\ 
\textsf{ivel$^2$} & 0.0133 & 0.0540 & 0.0980 & 0.3220\\ 
\textsf{ivel*date} & 0.0012 & 0.0250 & 0.0645 & 0.0527\\ 
\bottomrule
\end{tabular}
\end{table}

Using each of the first three columns displayed in Table \ref{tab:MPIPsMallard} one can extract a median probability model (MPM).  Following the same approach, with the last column in Table \ref{tab:MPIPsMallard}, we obtain the 50\% threshold model using the AIC weights. The MPM matches for RPE and FPE, and this model in turn is similar to that from EPE, but the latter excludes the \textsf{forest} term in the presence component.  In spite of this discrepancy, it is noteworthy that the MPIP using EPE for this term is 0.4305, being relatively close to the 0.5 threshold for the MPM.  The comparable model obtained using AIC weights is considerably larger than all the MPMs resulting with EPE, RPE and FPE, all of which are nested within it.

\begin{table}[!h]\caption{MPMs obtained from MPIPs using EPE, RPE and FPE and pseudo-MPM with AIC weights for the mallard dataset}\label{tab:MPMsMallard}
\begin{tabular}{l  l  l}
\toprule
& Detection & Presence\\
\midrule
{\bf EPE}&$\lrb{\1}$&$\lrb{\1,\textsf{elev},\textsf{forest}}$\\
{\bf RPE}&$\lrb{\1}$&$\lrb{\1,\textsf{elev},\textsf{forest},\textsf{length}}$\\
{\bf FPE}&$\lrb{\1}$&$\lrb{\1,\textsf{elev},\textsf{forest},\textsf{length}}$\\
{\bf AIC$_w$}&$\lrb{\1, \textsf{date}}$&$\lrb{\1,\textsf{elev},\textsf{forest},\textsf{length},\textsf{length*forest},\textsf{elev*length}}$\\
\bottomrule
\end{tabular}
\end{table}

Finally, Table \ref{tab:TopMallard} displays the five highest probability models (HPMs) under the three calculation alternatives, as well as those resulting from AIC based ranking.  Remarkably, the highest probability model is the same under the true posterior probabilities and the two estimation methods considered. Among the set of top models resulting from EPE, four are among the top five from RPE, and three are among those from FPE.  Additionally, the model ranked fifth using EPE, which does not match with any of the top five HPMs from RPE or FPE, is ranked eighth and ninth with RPE and FPE, respectively.  Also, models ranked fifth under RPE (which coincides with model four with FPE) and fifth under FPE, which are not among the top five with EPE, are respectively ranked eighth and seventh with EPE.  Again, more complex top models result from AIC selection in the presence components, and notably the model posterior probabilities are highly diluted across the model space, with the five top models concentrating only about 8\% of the posterior mass.  This contrasts markedly with the mass harnessed by the top five models with the other three methods, which are approximately 26\% with FPE, 43\% for RPE and 55\% with EPE.

\begin{table}[ht]\caption{Top five models with EPE, RPE, FPE and AIC for the mallard dataset}\label{tab:TopMallard}
\centering
\begin{tabular}{rllc}
EPE &  &  & \\ 
  \toprule
 & Detection & Presence & $p(M_y,M_z|\by)$ \\ 
  \midrule
1 &  $\lrb{\1}$ & $\{\1\textsf{,elev,forest}\}$ & 0.3101 \\ 
  2 &  $\lrb{\1}$ & \{$\1$\textsf{,elev,length,forest}\} & 0.0954 \\ 
  3 &  $\lrb{\1}$ & \{$\1$\textsf{,elev,length,forest,elev*length,length*forest}\} & 0.0634 \\ 
  4 &  $\lrb{\1}$ & \{$\1$\textsf{,elev,length,forest,elev*length}\} & 0.0420 \\ 
  5 &  $\lrb{\1}$ & \{$\1$\textsf{,elev,forest,elev*forest}\} & 0.0373 \\ 
   \bottomrule
 &  &  & \\ 
RPE &  &  & \\ 
\toprule
 & Detection & Presence & $p(M_y,M_z|\by)$ \\ 
\midrule
1 & $\lrb{\1}$ & $\{\1\textsf{,elev,forest}\}$ & 0.1821 \\ 
  2 & $\lrb{\1}$ & $\{\1\textsf{,elev,length,forest,elev*length,length*forest}\}$ & 0.0933 \\ 
  3 & $\lrb{\1}$ & $\{\1\textsf{,elev,length,forest,elev*length}\}$ & 0.0576 \\ 
  4 & $\lrb{\1}$ & $\{\1\textsf{,elev,length,forest}\}$ & 0.0572 \\ 
  5 & $\lrb{\1}$ & $\{\1\textsf{,elev,length,forest,length*forest}\}$ & 0.0431 \\ 
\bottomrule
 &  &  & \\ 
FPE &  &  & \\ 
 \toprule
 & Detection & Presence & $p(M_y,M_z|\by)$ \\ 
  \midrule
1 & $\lrb{\1}$ & \{$\1$\textsf{,elev,forest}\} & 0.1063 \\ 
  2 & $\lrb{\1}$ & \{$\1$\textsf{,elev,length,forest,elev*length,length*forest}\} & 0.0600 \\ 
  3 & $\lrb{\1}$ & \{$\1$\textsf{,elev,length,forest,elev*length}\} & 0.0354 \\ 
  4 & $\lrb{\1}$ & \{$\1$\textsf{,elev,length,forest,length*forest}\} & 0.0300 \\ 
  5 & \{$\1$,date\} & \{$\1$\textsf{,elev,forest}\} & 0.0284 \\ 
     \bottomrule
 &  &  & \\ 
AIC$_w$ &  &  & \\ 
  \toprule
 & Detection & Presence & AIC$_w(M_y,M_z|\by)$ \\ 
  \midrule
1 & \{$\1$,date\}  & \{$\1$\textsf{,elev,forest,length,elev*length,forest*length}\} & 0.0192 \\ 
  2 & \{$\1$,date\}  & \{$\1$\textsf{,elev,forest,length,length$^2$,elev*length,forest*length}\} & 0.0190 \\ 
  3 & \{$\1$\}  & \{$\1$\textsf{,elev,forest,length,length$^2$,elev*length,forest*length}\} & 0.0136 \\ 
  4 &  \{$\1$\} & \{$\1$\textsf{,elev,forest,length,elev*length,forest*length}\} & 0.0136 \\ 
  5 & \{$\1$,date$^2$\}  & \{$\1$\textsf{,elev,forest,length,elev*length,forest*length}\} & 0.0121 \\ 
   \bottomrule
\end{tabular}
\end{table}

The results in Tables \ref{tab:MPIPsMallard}-\ref{tab:TopMallard} indicate that estimating the model posterior probabilities using either RPE or FPE yield reasonable approximations to the actual posterior probabilities. In particular, all methods rank models similarly, and if model averaging was to be performed, these would all produce comparable results, as the derived MPIPs resemble each other under the three alternatives.  Nonetheless, following the results from Table \ref{tab:MPIPsMallard} we prefer RPEs, as these appear to be converging faster towards the benchmark posterior values (EPEs). These results are consistent with the findings from exhaustive simulation experiments conducted in \citet{Taylor-Rodriguez2015}, where overwhelming evidence was found in favor of renormalized model posterior estimates when compared to the frequency-based ones in the multiple linear regression problem.  For occupancy models, this behavior is more conspicuous in the detection component than in the presence one, possibly due to the additional uncertainty arising from only partially observing the presence indicators.  In addition to the observation that the renormalized posteriors are closer to those from enumeration, in larger model spaces where not all models are visited by the stochastic search, it is possible to calculate renormalized posteriors for a larger set of models than those visited, while with frequency-based estimates this is not possible.

\subsection{Blue hawker data}\label{subsec:bluehawk}

During 1999 and 2000, an intensive volunteer surveying effort coordinated by the Centre Suisse de Cartographie de la Faune (CSCF) was conducted to analyze the distribution of the blue hawker, {\it Ashna cyanea} (Odonata, Aeshnidae), a common dragonfly  in Switzerland.  Repeated visits to 1-ha pixels took place to obtain the corresponding detection history.  In addition to the survey outcome, the x- and y-coordinates, thermal level, the date of the survey, and the elevation were recorded.  Surveys were restricted to the known flight period of the blue hawker, which occurs between May 1 and October 10.  In total, 2,572 sites were surveyed at least once during the surveying period.  The number of surveys per site ranges from 1 to 22 times within each survey year, with as many as 67\% of the sites being surveyed only once, and only 5\% of the sites being surveyed more than 3 times.  As such, the analysis of this data set is an illustration of a considerably more challenging problem. 

 \citet{KG_10} summarize the results of this effort using AIC-based model comparisons. To select the predictors in the detection component, the authors follow a backwards elimination approach while keeping the presence component fixed at the most complex model. To select the presence model, they choose among a group of three models while using the chosen detection model. The full models  considered in this study are
\bea
\Phi^{-1}(p)&=&\lambda_0+\lambda_1 \textsf{year}+\lambda_2 \textsf{elev}+\lambda_3 \textsf{elev}^{2}+\lambda_4 \textsf{elev}^{3}+\lambda_5 \textsf{date}+\lambda_6 \textsf{date}^2\nonumber\\
\Phi^{-1}(\psi)&=&\alpha_0+\alpha_1\,\textsf{year}+\alpha_2\, \textsf{elev}+\alpha_3\, \textsf{elev}^{2}+\alpha_4\, \textsf{elev}^3,\nonumber\eea
where the term $\textsf{year}$ denotes $\mathcal{I}_{\lrb{\textsf{year}=2000}}$.

Assuming these full models and intercept only base models (and disregarding the polynomial hierarchy among predictors), the model space for this problem contains $2^{6+4}=1,024$ models in the joint model space. However, if the polynomial structure is respected, without considering interactions (for compatibility with the analysis in \citet{KG_10}), the size of the model space for the detection component reduces to 24 models, and to eight models for the presence. This corresponds to a total of 192 models in the combined space. In the exercise below, when using the proposed approach we enforce the strong heredity condition through the priors over the model space.  

As in the analysis of the Mallard dataset, we obtain the EPEs, the RPEs, and the FPEs.  The model ranks obtained with the posterior probabilities (or their estimates) are compared to those resulting from AIC selection.   The functions used to conduct selection with AIC did not constrain the model space to respect strong heredity, hence for the AIC selection all 1024 models were considered.  All results are compared to the models ultimately recommended by \citet{KG_10}, given by
\bea
\text{Detection:}&&\lrb{\1,\textsf{elev},\textsf{elev}^{2},\textsf{date}, \textsf{date}^2}\nonumber\\
\text{Presence:}&&\lrb{\1,\textsf{elev},\textsf{elev}^{2},\textsf{elev}^3}\nonumber.
\eea

\subsubsection{Results}

Table \ref{tab:MPMsBlue} shows the MPMs from either of the approaches considered obtained with the MPIPs found in Table \ref{tab:MPIPsBlue} of Appendix \ref{sec:appB}. The MPMs obtained with RPE and FPE coincide, and are similar to that from EPE, with the the latter additionally including the \textsf{elev}$^2$ term.  The pseudo-MPM that results when using AIC weights contains all the term included in the MPMs from RPE and FPE, but adds the \textsf{elev}$^3$ and \textsf{year} terms in the detection component.  Note that this model does not respect the polynomial hierarchy, including \textsf{elev}$^3$ but not \textsf{elev}$^2$.

\begin{table}[!h]\caption{MPMs obtained from MPIPs using EPE, RPE and FPE and pseudo-MPM with AIC weights for the blue hawker dataset}\label{tab:MPMsBlue}
\begin{tabular}{l  l  l}
\toprule
& Detection & Presence\\
\midrule
{\bf EPE}&\{\textsf{$\1$,date,date$^2$,elev,elev$^2$}\}&\{\textsf{$\1$,elev,elev$^2$}\}\\
{\bf RPE}&\{\textsf{$\1$,date,date$^2$,elev}\}&\{\textsf{$\1$,elev,elev$^2$}\}\\
{\bf FPE}&\{\textsf{$\1$,date,date$^2$,elev}\}&\{\textsf{$\1$,elev,elev$^2$}\}\\
{\bf AIC$_w$}&\{\textsf{$\1$,date,date$^2$,elev,elev$^3$,year}\}&\{\textsf{$\1$,elev,elev$^3$}\}\\
\bottomrule
\end{tabular}
\end{table}

The top ranked models in terms of the true (EPE) and estimated posterior probabilities (RPE and FPE), and from AIC-based selection are displayed in Table \ref{tab:HPMBlueHawk}.  The top model obtained with EPE, RPE and FPE are the same for both the presence and detection components, with the top AIC model not respecting the polynomial hierarchy in the detection component (including the \textsf{elev}$^3$ but not \textsf{elev}$^2$) and having only the \textsf{year} term in the presence component.  Interestingly, four out of the top five models found by EPE coincide with those from RPE, whereas only two from EPE are among the top 5 discovered with FPE, indicating again faster convergence of the renormalized estimates when compared to the frequency based ones.  Again, it is worth emphasizing that the probability mass with AIC weight is much more diluted across the model space than with any of its Bayesian counterparts.

\begin{table}[ht]\caption{Top ranked models using EPE, RPE, FPE and AIC weights for the blue hawker dataset}\label{tab:HPMBlueHawk}
\begin{tabular}{l  l  l c}
\toprule
& Detection & Presence & $p(M_y,M_z|\by)$\\
\midrule
{\bf EPE}&\{\textsf{$\1$,date,date$^2$,elev}\} &\{\textsf{$\1$,elev,elev$^2$}\} & 0.2090\\
{\bf RPE}&\{\textsf{$\1$,date,date$^2$,elev}\}&\{\textsf{$\1$,elev,elev$^2$}\} & 0.3725\\
{\bf FPE}&\{\textsf{$\1$,,date,date$^2$,elev}\}&\{\textsf{$\1$,elev,elev$^2$}\} & 0.1974\\
{\bf AIC$_w$}&\{\textsf{$\1$,date,date$^2$,elev,elev$^3$,year}\}& \{\textsf{$\1$,year}\} & 0.0422\\
\bottomrule
\end{tabular}
\end{table}

To conclude, a notable advantage of the Bayesian approach is that  the uncertainty associated to the choice of a particular model can be assessed using the model posterior probabilities, whereas this is not the case with AIC selection, as the AIC weights do not correspond to actual posterior probabilities.

\section{Discussion}\label{sec:discussion}

This paper developed the first objective Bayes methodology for variable selection using single-season site occupancy models, based on intrinsic priors derived from non-informative priors.  This solution uses latent variables to data-augment the analysis, helping to seamlessly calculate the model posterior probabilities. Working on the latent scale additionally facilitates the construction of a straightforward MCMC sampler and posterior estimation using sample averages.

Because the intrinsic priors are built from non-informative priors, the need for hyperparameter specification is avoided, making the method entirely automatic and widely applicable. Additionally, the types of prior distributions assumed on the model space (HIP, HOP and HUP) enforce the heredity constraints required when performing selection with interactions and higher-order polynomial predictors. These classes also allow for stronger penalization than the usual equal probability prior, further helping control the false positive rate. These have been shown to be particularly useful in problems with small and moderate sample sizes \citep[for more details see][]{Taylor-Rodriguez2015}. % Being fully Bayesian, our methodology allows conducting inference under uncertainty and under model selection \citep[see][for a detailed description]{Womack2014}. 
An important advantage of our method, relative to the AIC-based selection, is that the resulting model posterior probabilities provide a measure of uncertainty associated with choosing a particular model.  

The stochastic search algorithm can be used to thoroughly explore large model spaces using the renormalized posterior estimates (instead of the frequency-based ones). This tool will allow practitioners to explore the model space without having to enumerate it or preselect a subset of models, enabling its use with larger model spaces.

The simulation experiments confirmed the ability of the method to identify the predictors present in the true model when considering both the highest and median probability models.   The objective Bayes method proved to be competitive with AIC in detecting true predictors, and greatly outperformed AIC in reducing the number of false positive predictors included in the models with high posterior probabilities.

The software used throughout the article was built into the R package \textsf{OccOBayes} available at request. This package includes functions to run the variable selection procedure, as well as some auxiliary functions to validate a set of ``best'' models using a hold-out data set.

\appendix
%\section{Appendix}
\section{Model Selection Algorithm}
%Having the parameter intrinsic priors in place and knowing the form of the model posterior probabilities, it is finally possible to develop a strategy to conduct model selection for the occupancy framework.

For each of the two components of the model --presence and detection-- the algorithm first draws the set of active predictors  (i.e., $A_z$ and $A_y$) together with their corresponding parameters. This is a reversible jump step that uses a Metropolis Hastings correction with proposal distributions given by
\bea
q(A_z^{*}|\bz_o,\bz_u^{(t)},\bv^{(t)},M_{A_z})&=&\frac{1}{2}\left(p(M_{A_z^{*}}|\bz_o,\bz_u^{(t)},\bv^{(t)},\cM_z,M_{A_z^{*}}\in L(M_{A_z}) )+\frac{1}{|L(M_{A_z})|}\right)\nonumber\\
q(A_y^{*}|\by,\bz_o,\bz_u^{(t)},\bw^{(t)},M_{A_y})&=&\frac{1}{2}\left(p(M_{A_w^{*}}|\by,\bz_o,\bz_u^{(t)},\bw^{(t)},\cM_y,M_{A_y^{*}}\in L(M_{A_y}) )+\frac{1}{|L(M_{A_y})|}\right),\nonumber\\
&&
\eea
where $L(M_{A_z})$ and $L(M_{A_y})$ denote the sets of models obtained by adding or removing one predictor at a time from the corresponding feasible sets of nodes in $M_{A_z}$ and $M_{A_y}$, respectively. Here $\bz_0$ are the observed presence indicators and $\bz_u$ are those that are unobserved.

To promote mixing, this step is followed by an additional draw from the full conditionals of $\balpha$ and $\blambda$. The densities $p(\balpha_0|.)$, $p(\balpha_{A}|.)$, $p(\blambda_0|.)$, and $p(\blambda_{A}|.)$ can be sampled from exactly via Gibbs steps. Using the notation $a|\cdot$ to denote the random variable $a$ conditioned on all other parameters and on the data, these densities are given by
\begin{itemize}
\item $\balpha_0|\cdot \sim \mathcal{N}\lrp{(\bX_0^{\prime}\bX_0)^{-1}\bX_0^{\prime}\bv, (\bX_0^{\prime}\bX_0)^{-1}}$,

\item $\balpha_{A}|\cdot \sim \mathcal{N}\lrp{\bmu_{\balpha_{A}}, \Sigma_{\balpha_{A}} }$,  where the mean vector and the covariance matrix are given by $\Sigma_{\balpha_{A}}=\frac{2N}{2N+p_{A_z}}(\bX_{A}^{\prime}\bX_{A})^{-1}$ and $\bmu_{\balpha_{A}}=\lrp{\Sigma_{\balpha_{A}}\bX_{A}^{\prime}\bv}$,

\item $\blambda_0|\cdot \sim \mathcal{N}\lrp{(\bQ_0^{\prime}\bQ_0)^{-1}\bQ_0^{\prime}\bw, (\bQ_0^{\prime}\bQ_0)^{-1}}$, and

\item $\blambda_{A}|\cdot \sim \mathcal{N}\lrp{\bmu_{\blambda_{A}}, \Sigma_{\blambda_{A}} }$, analogously with mean and covariance matrix given by $\Sigma_{\blambda_{A}}=\frac{2\jd}{2\jd+p_{A_y}}(\bQ_{A}^{\prime}\bQ_{A})^{-1}$ and $\bmu_{\blambda_{A}}=\lrp{\Sigma_{\blambda_{A}} \bQ_{A}^{\prime}\bw}$.
\end{itemize}

Finally, Gibbs sampling steps are also available for the unobserved occupancy indicators $\bz_u$, and for the corresponding latent variables $\bv$ and $\bw$.  The full conditional posterior densities for $\bz_{u}^{(t+1)}$, $\bv^{(t+1)}$, and $\bw^{(t+1)}$ are those described in \citet{DTR_12} for the single-season probit model.

The following steps summarize the stochastic search algorithm:
\begin{enumerate}
\item Initialize $A_y^{(0)}, A_z^{(0)}, \bz_{u}^{(0)}, \bv^{(0)}, \bw^{(0)},\balpha_0^{(0)}, \blambda_{0}^{(0)}$.%\balpha_{r,A}^{(0)}, , \blambda_{r,A}^{(0)}$.

\item Sample the model indices and corresponding parameters:
%%%%%%%%% sampling for occupancy process
\begin{enumerate}
\item Draw simultaneously
\begin{itemize}
\item $A_z^*\sim q(A_z|\bz_o,\bz_u^{(t)},\bv^{(t)},M_{A_z})$,
\item $\balpha_{0}^*\sim p(\balpha_{0}|\bv^{(t)})$, and
\item $\balpha_{A^*}^*\sim p(\balpha_{A}|M_{A_z^*},\bv^{(t)})$.
\end{itemize}

\item Accept $(M_{A_z}^{(t+1)},\balpha_{0}^{(t+1),1},\balpha_{A}^{(t+1),1})=(M_{A_z^*},\balpha_{0}^{*},\balpha_{A^*}^*)$ with probability
\begin{eqnarray*}
\delta_z&=&\min{\lrp{1,\frac{p(M_{A_z^*}|\bz_o,\bz_u^{(t)},\bv^{(t)})}{p(M_{A_z^{(t)} }|\bz_o,\bz_u^{(t)},\bv^{(t)})} \frac{q(A_z^{(t)}|\bz_o,\bz_u^{(t)},\bv^{(t)},M_{A_z^*})}{q(A_z^*|\bz_o,\bz_u^{(t)},\bv^{(t)},M_{A_z})}} },
\end{eqnarray*}
otherwise, let $(M_{A_z}^{(t+1)},\balpha_{0}^{(t+1),1},\balpha_{A}^{(t+1),1})=(A_z^{(t)}, \balpha_{0}^{(t),2},\balpha_{A}^{(t),2})$.

\item Sample simultaneously
\begin{itemize}
\item $A_y^*\sim q(A_y|\by,\bz_o,\bz_u^{(t)},\bw^{(t)},M_{A_y})$,
\item $\blambda_{0}^*\sim p(\blambda_{0}|\bw^{(t)})$,  and
\item $\blambda_{A^*}^*\sim p(\blambda_{A}|M_{A_y^*},\bw^{(t)})$.
\end{itemize}

\item  Accept $(M_{A_y}^{(t+1)},\blambda_{0}^{(t+1),1},\blambda_{A}^{(t+1),1})=(M_{A_y^*},\blambda_{0}^*,\blambda_{A^*}^*)$ with probability
\begin{eqnarray*}
\delta_y&=&\min{\lrp{1,\frac{p(M_{A_z^*}|\by,\bz_o,\bz_u^{(t)},\bw^{(t)})}{p(M_{A_z^{(t)} }|\by,\bz_o,\bz_u^{(t)},\bw^{(t)})} \frac{q(A_z^{(t)}|\by,\bz_o,\bz_u^{(t)},\bw^{(t)},M_{A_y^*})}{q(A_z^*|\by,\bz_o,\bz_u^{(t)},\bw^{(t)},M_{A_y})}} },
\end{eqnarray*}
otherwise, let $(M_{A_y}^{(t+1)},\blambda_{0}^{(t+1),1},\blambda_{A}^{(t+1),1})=(A_y^{(t)},\blambda_{0}^{(t),2},\blambda_{A}^{(t),2})$.

\end{enumerate}

%%%%%%%%% sampling for occupancy process
\item Sample base model parameters:

\begin{enumerate}

\item Draw $\balpha_0^{(t+1),2}\sim p(\balpha_0|\bv^{(t)})$.
\item Draw $\lambda_0^{(t+1),2}\sim p(\blambda_0|\bw^{(t)})$.

\end{enumerate}

%%%%%%%%% sampling model parameters
\item To improve mixing, resample model coefficients not present the base model but are in $M_A$:

\begin{enumerate}
\item Draw $\balpha_{A}^{(t+1),2}\sim p(\balpha_{A}|M_{A_z^{(t+1)}},\bv^{(t)})$.

\item Draw $\blambda_{A}^{(t+1),2}\sim p(\blambda_{A}|M_{A_y^{(t+1)}},\bw^{(t)})$.

\end{enumerate}

%%%%%%%%% sampling for occupancy process
\item Sample latent and missing (unobserved) variables:

\begin{enumerate}

\item Sample $\bz_u^{(t+1)}\sim p(\bz_u|M_{A_z^{(t+1)}},\by,\balpha_{A}^{(t+1),2},\balpha_0^{(t+1),2},\blambda_{A}^{(t+1),2},\blambda_0^{(t+1),2} )$

\item Sample $\bv^{(t+1)}\sim p(\bv|M_{A_z^{(t+1)}},\bz_o,\bz_u^{(t+1)},\balpha_{A}^{(t+1),2},\balpha_0^{(t+1),2}  )$

\item Sample $\bw^{(t+1)}\sim p(\bw|M_{A_y^{(t+1)}},\bz_o,\bz_u^{(t+1)},\blambda_{A}^{(t+1),2},\blambda_0^{(t+1),2}  )$

\end{enumerate}

\end{enumerate}

\section{Additional tables blue hawker analysis}\label{sec:appB}

\begin{table}[!ht]
\caption{MPIPs with EPE, RPE and FPE and AIC weights, obtained from joint model for presence (top) and detection (bottom) components  for the blue hawker dataset}\label{tab:MPIPsBlue}
\centering
\begin{tabular}{lcccc}
  \toprule
 & EPE & RPE & FPE & AIC$_w$\\ 
  \midrule
\textsf{elev} & 0.5346 & 0.7971 & 0.8338 & 0.5538\\ 
\textsf{elev$^2$} &  0.5228 & 0.7885 & 0.7956 & 0.3542\\ 
\textsf{elev$^3$} & 0.2041 & 0.2923 & 0.2988 & 0.6626\\ 
\textsf{year} &  0.1130 & 0.0676 & 0.3421 & 0.4125\\ 
\bottomrule
 & & & &\\ 
  \toprule
 & EPE & RPE & FPE & AIC$_w$\\ 
  \midrule
\textsf{date} & 0.9999 & 1.0000 & 1.0000 & 1.0000\\ 
\textsf{date$^2$} & 0.9999 & 0.9737 & 0.9632 & 1.0000 \\ 
\textsf{elev} & 0.9852 & 0.9590 & 0.9582 & 0.9630\\ 
\textsf{elev$^2$} & 0.5170 & 0.2591 & 0.2797 &  0.4099\\ 
\textsf{elev$^3$} & 0.2601 & 0.0921 & 0.1114 & 0.5453\\ 
\textsf{year} & 0.2169 & 0.0658 & 0.2845 & 0.5566\\ 
   \bottomrule
\end{tabular}
\end{table}

\begin{table}[ht]\caption{Top five models with RPE, FPE and AIC for the blue hawker dataset}\label{tab:Top5BlueHawk}
\centering
\begin{tabular}{rllc}
EPE & & & \\
  \toprule
 & Detection & Presence & Post \\ 
  \midrule
1 & \{\textsf{$\1$, elev,date,date$^2$}\} & \{\textsf{$\1$, elev,elev$^2$}\} & 0.2090 \\ 
  2 & \{\textsf{$\1$, elev,date,elev$^2$,date$^2$,elev$^3$}\} & \{\textsf{$\1$}\} & 0.1763 \\ 
  3 & \{\textsf{$\1$, elev,date,elev$^2$,date$^2$}\} & \{\textsf{$\1$}\} & 0.1747 \\ 
  4 & \{\textsf{$\1$, elev,date,date$^2$}\} & \{\textsf{$\1$, elev,elev$^2$,elev$^3$}\} & 0.1200 \\ 
  5 & \{\textsf{$\1$, year,elev,date,elev$^2$,date$^2$,elev$^3$}\} & \{\textsf{$\1$}\} & 0.0568 \\ 
   \bottomrule
   &  &  & \\  
RPE &  &  & \\
 \toprule
 & Detection & Presence & $p(M_y,M_z|\by)$ \\ 
\midrule
1 & \{\textsf{$\1$,elev,date,date$^2$}\} & \{\textsf{$\1$,elev,elev$^2$}\} & 0.3725 \\ 
  2 & \{\textsf{$\1$,elev,date,date$^2$}\} & \{\textsf{$\1$,elev,elev$^2$,elev$^3$}\} & 0.2058 \\ 
  3 & \{\textsf{$\1$,elev,date,elev$^2$,date$^2$}\} & \{\textsf{$\1$}\} & 0.1055 \\ 
  4 & \{\textsf{$\1$,elev,date,elev$^2$,date$^2$,elev$^3$}\} & \{\textsf{$\1$}\} & 0.0656 \\ 
  5 & \{\textsf{$\1$,elev,date,date$^2$}\} & \{\textsf{$\1$,year,elev,elev$^2$}\} & 0.0308 \\ 
    \bottomrule
 &  &  & \\ 
 FPE &  &  & \\
\toprule
 & Detection & Presence & $p(M_y,M_z|\by)$ \\ 
  \midrule
1 & \{\textsf{$\1$,elev,date,date$^2$}\} & \{\textsf{$\1$,elev,elev$^2$}\} & 0.1974 \\ 
  2 & \{\textsf{$\1$,elev,date,date$^2$}\} & \{\textsf{$\1$,elev,elev$^2$,elev$^3$}\} & 0.1138 \\ 
  3 & \{\textsf{$\1$,elev,date,date$^2$}\} & \{\textsf{$\1$,year,elev,elev$^2$}\} & 0.1023 \\ 
  4 & \{\textsf{$\1$,year,elev,date,date$^2$}\} & \{\textsf{$\1$,elev,elev$^2$}\} & 0.0728 \\ 
  5 & \{\textsf{$\1$,elev,date,date$^2$}\} & \{\textsf{$\1$,year,elev,elev$^2$,elev$^3$}\} & 0.0599 \\
     \bottomrule
      &  &  & \\ 
 AIC$_w$ &  &  & \\
 \toprule
 & Detection & Presence & AIC$_w(M_y,M_z|\by)$ \\ 
  \midrule
1 & \{\textsf{$\1$,date,date$^2$,elev,elev$^3$,year}\} & \{\textsf{$\1$,year}\} & 0.0422 \\ 
  2 & \{\textsf{$\1$,date,date$^2$,elev}\} & \{\textsf{$\1$,elev,elev$^3$}\} & 0.0403 \\ 
  3 & \{\textsf{$\1$,date,date$^2$,elev,year}\} & \{\textsf{$\1$,elev,elev$^3$}\} & 0.0348 \\ 
  4 & \{\textsf{$\1$,date,date$^2$,elev,year}\} & \{\textsf{$\1$,elev,elev$^3$,year}\} & 0.0321 \\ 
  5 & \{\textsf{$\1$,date,date$^2$,elev,elev$^3$,year}\} &\{\textsf{$\1$}\}  & 0.0254 \\ 
   \bottomrule
\end{tabular}
\end{table}

\clearpage 

\bibliographystyle{apalike}%\bibliographystyle{plainnat}
\bibliography{BibDissertation}

\begin{thebibliography}{}

\bibitem[Akaike, 1983]{A_73}
Akaike, H. (1983).
\newblock Information measures and model selection.
\newblock {\em Bull. Int. Statist. Inst.}, 50:277--290.

\bibitem[Albert and Chib, 1993]{ACh_93}
Albert, J.~H. and Chib, S. (1993).
\newblock {Bayesian-analysis of binary and polychotomous response data}.
\newblock {\em Journal of the American Statistical Association},
  88(422):669--679.

\bibitem[Berger and Pericchi, 1996]{Berger1996}
Berger, J. and Pericchi, L. (1996).
\newblock {The intrinsic Bayes factor for model selection and prediction}.
\newblock {\em Journal of the American Statistical Association},
  91(433):109--122.

\bibitem[Burnham and Anderson, 2003]{burnham2003model}
Burnham, K. and Anderson, D. (2003).
\newblock {\em Model Selection and Multimodel Inference: A Practical
  Information-Theoretic Approach}.
\newblock Springer New York.

\bibitem[Burnham, 2004]{Burnham2004}
Burnham, K.~P. (2004).
\newblock {Multimodel Inference: Understanding AIC and BIC in Model Selection}.
\newblock {\em Sociological Methods {\&} Research}, 33(2):261--304.

\bibitem[Casella and Moreno, 2006]{Casella2006}
Casella, G. and Moreno, E. (2006).
\newblock {Objective Bayesian Variable Selection}.
\newblock {\em Journal of the American Statistical Association},
  101(473):157--167.

\bibitem[Chib, 1995]{chib1995marginal}
Chib, S. (1995).
\newblock Marginal likelihood from the gibbs output.
\newblock {\em Journal of the American Statistical Association},
  90(432):1313--1321.

\bibitem[Chib and Jeliazkov, 2001]{chib2001marginal}
Chib, S. and Jeliazkov, I. (2001).
\newblock Marginal likelihood from the metropolis--hastings output.
\newblock {\em Journal of the American Statistical Association},
  96(453):270--281.

\bibitem[Chipman, 1996]{Chipman1996}
Chipman, H. (1996).
\newblock {Bayesian variable selection with related predictors}.
\newblock {\em Canadian Journal of Statistics}, 24(1):17--36.

\bibitem[Dorazio and Taylor-Rodriguez, 2012]{DTR_12}
Dorazio, R. and Taylor-Rodriguez, D. (2012).
\newblock {A Gibbs sampler for Bayesian analysis of site-occupancy data}.
\newblock {\em Methods in Ecology and Evolution}.

\bibitem[Fiske and Chandler, 2011]{FiskeUNM}
Fiske, I. and Chandler, R. (2011).
\newblock {unmarked: An R package for fitting hierarchical models of wildlife
  occurrence and abundance}.
\newblock {\em Journal of Statistical Software}, 43(10).

\bibitem[Flegal and Gong, 2013]{flegal2013relative}
Flegal, J.~M. and Gong, L. (2013).
\newblock Relative fixed-width stopping rules for markov chain monte carlo
  simulations.
\newblock {\em arXiv preprint arXiv:1303.0238}.

\bibitem[Green, 1995]{Green1995}
Green, P.~J. (1995).
\newblock {Reversible Jump Markov Chain Monte Carlo Computation and Bayesian
  Model Determination}.
\newblock {\em Biometrika}, 82(4):711--732.

\bibitem[Guillera-Arroita et~al., 2014]{Guillera-Arroita2014}
Guillera-Arroita, G., Lahoz-Monfort, J.~J., MacKenzie, D.~I., Wintle, B.~A.,
  and McCarthy, M.~A. (2014).
\newblock {Ignoring Imperfect Detection in Biological Surveys Is Dangerous: A
  Response to: Fitting and Interpreting Occupancy Models'}.
\newblock {\em PLoS ONE}, 9(7):e99571.

\bibitem[Hooten and Hobbs, 2015]{Hooten2015}
Hooten, M.~B. and Hobbs, N.~T. (2015).
\newblock {A guide to Bayesian model selection for ecologists}.
\newblock {\em Ecological Monographs}, 85(1):3--28.

\bibitem[Hurvich and Tsai, 1989]{Hurvich1989}
Hurvich, C.~M. and Tsai, C.-L. (1989).
\newblock {Regression and time series model selection in small samples}.
\newblock {\em Biometrika}, 76:297--307.

\bibitem[Kass and Wasserman, 1996]{Kass1996}
Kass, R.~E. and Wasserman, L. (1996).
\newblock {The Selection of Prior Distributions by Formal Rules}.
\newblock {\em Journal of the American Statistical Association}, 91(435):1343.

\bibitem[K\'{e}ry et~al., 2010]{KG_10}
K\'{e}ry, M., Gardner, B., and Monnerat, C. (2010).
\newblock Predicting species distributions from checklist data using
  site-occupancy models.
\newblock {\em Journal of Biogeography}, 37(10):1851--1862.
\newblock K\'{e}ry, Marc Gardner, Beth Monnerat, Christian.

\bibitem[K{\'{e}}ry et~al., 2005]{Kery2005}
K{\'{e}}ry, M., Royle, J.~A., and Schmid, H. (2005).
\newblock {Modeling avian abundance from replicated counts using binomial
  mixture models}.
\newblock {\em Ecological Applications}, 15(4):1450--1461.

\bibitem[K{\'{e}}ry and Schmid, 2004]{Kery2004}
K{\'{e}}ry, M. and Schmid, H. (2004).
\newblock {Monitoring programs need to take into account imperfect species
  detectability}.
\newblock {\em Basic and Applied Ecology}, 5(1):65--73.

\bibitem[Leon-Novelo et~al., 2012]{Leon-Novelo2012}
Leon-Novelo, L., Moreno, E., and Casella, G. (2012).
\newblock {Objective Bayes model selection in probit models.}
\newblock {\em Statistics in medicine}, 31(4):353--65.

\bibitem[Liu and Wu, 1999]{liu1999parameter}
Liu, J.~S. and Wu, Y.~N. (1999).
\newblock Parameter expansion for data augmentation.
\newblock {\em Journal of the American Statistical Association},
  94(448):1264--1274.

\bibitem[MacKenzie et~al., 2002]{MacKetal_02}
MacKenzie, D.~I., Nichols, J.~D., Lachman, G.~B., Droege, S., Royle, J.~A., and
  Langtimm, C.~A. (2002).
\newblock Estimating site occupancy rates when detection probabilities are less
  than one.
\newblock {\em Ecology}, 83(8):2248--2255.

\bibitem[Mazerolle and Mazerolle, 2013]{Mazerolle2013}
Mazerolle, M. and Mazerolle, M. (2013).
\newblock {Package 'AICcmodavg'}.
\newblock (c).

\bibitem[McQuarrie et~al., 1997]{McQuarrie1997}
McQuarrie, A., Shumway, R., and Tsai, C.-L. (1997).
\newblock {The model selection criterion AICu}.
\newblock {\em Statistics \& Probability Letters}, 34(3):285--292.

\bibitem[Moreno et~al., 1998]{MBR_98}
Moreno, E., Bertolino, F., and Racugno, W. (1998).
\newblock An intrinsic limiting procedure for model selection and hypotheses
  testing.
\newblock {\em Journal of the American Statistical Association},
  93(444):1451--1460.

\bibitem[Peixoto, 1987]{P_87}
Peixoto, J.~L. (1987).
\newblock Hierarchical variable selection in polynomial regression models.
\newblock {\em American Statistician}, 41(4):311--313.

\bibitem[Peixoto, 1990]{P_90}
Peixoto, J.~L. (1990).
\newblock A property of well-formulated polynomial regression-models.
\newblock {\em American Statistician}, 44(1):26--30.

\bibitem[P\'{e}rez and Berger, 2002]{Perez2002}
P\'{e}rez, J. and Berger, J. (2002).
\newblock {Expected-posterior prior distributions for model selection}.
\newblock {\em Biometrika}, pages 491--511.

\bibitem[Rao and Wu, 2001]{rao2001}
Rao, C.~R. and Wu, Y. (2001).
\newblock {\em On model selection}, volume Volume 38 of {\em Lecture
  Notes--Monograph Series}, pages 1--57.
\newblock Institute of Mathematical Statistics, Beachwood, OH.

\bibitem[Roy and Hobert, 2007]{roy2007convergence}
Roy, V. and Hobert, J.~P. (2007).
\newblock Convergence rates and asymptotic standard errors for markov chain
  monte carlo algorithms for bayesian probit regression.
\newblock {\em Journal of the Royal Statistical Society: Series B (Statistical
  Methodology)}, 69(4):607--623.

\bibitem[Taylor-Rodriguez et~al., 2015]{Taylor-Rodriguez2015}
Taylor-Rodriguez, D., Womack, A., and Bliznyuk, N. (2015).
\newblock {Bayesian Variable Selection on Model Spaces Constrained by Heredity
  Conditions}.
\newblock {\em Journal of Computational and Graphical Statistics}, (July
  2015):0--0.

\bibitem[Wasserman, 2000]{Wasserman2000}
Wasserman, L. (2000).
\newblock {Bayesian Model Selection and Model Averaging.}
\newblock {\em Journal of mathematical psychology}, 44(1):92--107.

\bibitem[Yang, 2005]{Yang2005}
Yang, Y. (2005).
\newblock {Can the strengths of \textsc{AIC} and \textsc{BIC} be shared? A
  conflict between model identification and regression estimation}.
\newblock {\em Biometrika}, 92(4):937--950.

\end{thebibliography}

\bigskip

{\bf Acknowledgments} $\quad$ The first three authors were supported by the National Science Foundation grant DMS-1105127. % (PIs George Casella and Linda J.~Young).
Taylor-Rodriguez was additionally supported by the National Science Foundation under grant DMS-1127914 to the Statistical and Applied Mathematical Sciences Institute. Bliznyuk was additionally supported by the National Institutes of Health grants U54GM111274 and R21AI119773. Any opinions, findings, and conclusions or recommendations expressed in this material are those of the authors and do not necessarily reflect the views of the National Science Foundation or the National Institutes of Health. The blue hawker dataset was kindly provided by Christian Monnerat and Marc K\'{e}ry and authorized for use by the Swiss Biodiversity Monitoring program of the Swiss Federal Office for the Environment (FOEN).

\end{document}